\documentclass[a4paper,10pt,fleqn,usenames,dvipsnames]{article}

\usepackage{jheppub}

\usepackage{placeins}

\usepackage[OT1]{fontenc}
\usepackage{bold-extra}
\usepackage{amsmath}
\usepackage{amssymb}

\usepackage{url}
\usepackage{longtable}
\usepackage{multirow}
\usepackage{caption}
\usepackage[labelformat=simple]{subcaption}

\usepackage[nodayofweek]{datetime}
\usepackage{enumerate}
\usepackage{xspace}
\usepackage{xcolor}
\usepackage{graphicx}

\usepackage{float}
\usepackage[capitalise]{cleveref}

\newcommand{\drapes}{\mbox{\textsc{Drapes}}\xspace}
\newcommand{\drapesSR}{\mbox{\drapes~SR}\xspace}
\newcommand{\drapesSB}{\mbox{\drapes~SB}\xspace}
\newcommand{\drapesGen}{\mbox{\drapes~$\emptyset$}\xspace}
\newcommand{\drapesMC}{\mbox{\drapes~MC}\xspace}
\newcommand{\genSR}{\mbox{\textsc{GenSR}}\xspace}

\newcommand*\rfrac[2]{{}^{#1}\!/_{#2}}

\newcommand{\diff}{\mathop{}\!\mathrm{d}}

\newcommand{\pt}{\ensuremath{{p}_\mathrm{T}}\xspace}

\newcommand{\CURTAINs}{\mbox{\textsc{Curtain}s}\xspace}
\newcommand{\FfF}{\mbox{\textsc{Curtain}sF4F}\xspace}
\newcommand{\CATHODE}{\mbox{\textsc{Cathode}}\xspace}
\newcommand{\FETA}{\textsc{Feta}\xspace}
\newcommand{\CWoLa}{\mbox{\textsc{CWoLa}}\xspace}
\newcommand{\mjj}{\ensuremath{m_{JJ}}\xspace}

\title{Improving new physics searches with diffusion models for event observables and jet constituents}
\author{Debajyoti Sengupta,}
\author{Matthew Leigh,}
\author{John Andrew Raine,}
\author{Samuel Klein,}
\author{Tobias Golling}

\emailAdd{debajyoti.sengupta@unige.ch}
\emailAdd{matthew.leigh@unige.ch}
\emailAdd{john.raine@unige.ch}

\affiliation{Département de physique nucléaire et corpusculaire, Université de Genève, 1211 Genève, Switzerland}

\abstract{%
We introduce a new technique called \drapes to enhance the sensitivity in searches for new physics at the LHC.
By training diffusion models on side-band data, we show how background templates for the
signal region can be generated either directly from noise, or by partially applying the 
diffusion process to existing data.
In the partial diffusion case, data can be drawn from side-band regions, with the inverse diffusion performed for new target conditional values,
or from the signal region, preserving the distribution over the conditional property that defines the signal region.
We apply this technique to the hunt for resonances using the LHCO di-jet dataset, and achieve state-of-the-art performance for background template generation using high level input features.
We also show how \drapes can be applied to low level inputs with jet constituents, reducing the model dependence on the choice of input observables.
Using jet constituents we can further improve sensitivity to the signal process, but observe a loss in performance where the signal significance before applying any selection is below 4$\sigma$.
}

\begin{document}
    \maketitle

    \section{Introduction}
    The standard model of particle physics is one of the most successful scientific theories ever,
accurately describing the interactions of elementary particles for three of the fundamental forces of nature
in a quantum field theory.
However, it is known to be incomplete as it cannot describe gravity,
provide an explanation for dark matter, or address the (anti)matter imbalance in the universe, amongst other observed phenomena.
At the energy and intensity frontier, experiments at the Large Hadron Collider~\cite{Evans:2008zzb} at CERN,
such as the ATLAS and CMS experiments~\cite{ATLAS:2008xda,CMS:2008xjf},
hunt for new physics phenomena which could provide an insight into physics beyond the standard model~(BSM).

Dedicated analyses optimised for a wide variety of BSM models have been performed, but as yet no significant evidence of new physics has been uncovered~\cite{ATLAS:2021ilc,ATLAS:2021zqc,ATLAS:2021yfa,CMS:summary1,CMS:summary2,CMS:summary3}.
Recently, focus has turned to searches with broader sensitivity in the context of anomaly detection.
Instead of aiming for high sensitivity to a small subset of models,
these approaches target sensitivity to a much wider spectrum of possible new physics scenarios, at the expense of sensitivity to any one process~\cite{Kasieczka:2021xcg,Aarrestad:2021oeb}.

A cornerstone of these model agnostic searches is the bump hunt, which identifies localised excesses of data above a falling background, typically as a resonance in an invariant mass.
One of the drawbacks to the standard bump hunt is that the sensitivity is limited to the invariant mass after applying a selection to define a signal enriched sample.
To enhance the sensitivity, machine learning approaches which improve the sensitivity are studied,
in particular, weakly supervised classifiers (\CWoLa) trained to separate signal region data from a reference background sample~\cite{cwola,cwolabump}.
For this approach it is important that the reference background sample closely matches the true background data in the signal region,
otherwise the classifier can learn to separate events based on the background mismodelling, rather than differences between signal and background data.
Several machine learning methods have been developed for this utilising normalizing flows or multidimensional sample reweighting~\cite{anode,cathode,salad,curtains,curtainsf4f,feta,LaCathode}.
These methods, however, only form a subset of the array of anomaly detection techniques being developed in high energy physics~\cite{DAgnolo:2018cun,DAgnolo:2019vbw,Farina:2018fyg,Heimel:2018mkt,Roy:2019jae,Cerri:2018anq,Blance:2019ibf,Hajer:2018kqm,DeSimone:2018efk,Mullin:2019mmh,Dillon:2019cqt,Aguilar-Saavedra:2017rzt,Romao:2019dvs,Romao:2020ojy,Amram:2020ykb,Cheng:2020dal,Khosa:2020qrz,Thaprasop:2020mzp,Alexander:2020mbx,Mikuni:2020qds,vanBeekveld:2020txa,Park:2020pak,Faroughy:2020gas,Chakravarti:2021svb,Batson:2021agz,Blance:2021gcs,Bortolato:2021zic,Collins:2021nxn,Dillon:2021nxw,Finke:2021sdf,Shih:2021kbt,Atkinson:2021nlt,Kahn:2021drv,Dorigo:2021iyy,Caron:2021wmq,Govorkova:2021hqu,Kasieczka:2021tew,Volkovich:2021txe,Govorkova:2021utb,Ostdiek:2021bem,Fraser:2021lxm,Jawahar:2021vyu,Herrero-Garcia:2021goa,Aguilar-Saavedra:2021utu,Tombs:2021wae,Lester:2021aks,Mikuni:2021nwn,Chekanov:2021pus,dAgnolo:2021aun,Canelli:2021aps,Ngairangbam:2021yma,Bradshaw:2022qev,Aguilar-Saavedra:2022ejy,Buss:2022lxw,Alvi:2022fkk,Dillon:2022tmm,Birman:2022xzu,Letizia:2022xbe,Fanelli:2022xwl,Finke:2022lsu,Verheyen:2022tov,Dillon:2022mkq,Caron:2022wrw,Park:2022zov,Kamenik:2022qxs,Kasieczka:2022naq,Araz:2022zxk,Schuhmacher:2023pro,Roche:2023int,Vaslin:2023lig,ATLAS:2023azi,Chekanov:2023uot,CMSECAL:2023fvz,Bickendorf:2023nej,Freytsis:2023cjr,Metodiev:2023izu}.
% Several methods have been proposed to produce this tempalte with modern machine learning methods~\cite{anode,cathode,salad,curtains,feta,curtainsf4f}.

Recently, diffusion models~\cite{Song2020,Karras2022,yang2023diffusion} and continuous normalizing flows~\cite{chen2019neural_CNF,liu2022flow_flowmatching1, albergo2023building_flowmatching2, lipman2023flow_flowmatching3} have demonstrated exceptional performance in high energy physics applications~\cite{CaloScore,pcjedi,fpcd,Shmakov:2023kjj,Butter:2023fov,pcdroid,glam,Mikuni:2023tqg,Buhmann:2023kdg,epicjedi,Buhmann:2023acn,Heimel:2023mvw,Devlin:2023jzp,Heimel:2023ngj,Butter:2023ira,Birk:2023efj},
notably for reproducing the underlying kinematics of both calorimeter and particle showers.
These models have outperformed previous state-of-the-art approaches using normalizing flows and generative adversarial networks in both tasks.

In this work we introduce a new method, \drapes (Denoising resonant anomalies by perturbing existing samples), which leverages state of the art diffusion models to generate reference background samples in a fully data driven approach for resonant anomaly searches.
After training \drapes on side-band data, reference background samples can be generated in the signal region either by sampling from noise,
or by partially applying the forward and reverse diffusion processes on existing data drawn from either the side-bands, the signal region, or another reference sample.
We apply \drapes to the LHCO R\&D dataset~\cite{LHCOlympics} and demonstrate state-of-the art performance in training \CWoLa classifiers for a range of signal fractions.
Due to the flexibility of diffusion models, we can apply \drapes to both high level features and low level jet constituents.

This work represents one of the first applications of diffusion models\footnote{Here we can also include continuous normalizing flows. See Ref.~\cite{epicjedi} for a comparison with diffusion models, which introduces the umbrella term continuous time generative models.} for the task of sample generation in weakly supervised searches for new physics.
Concurrent with the development of this work a similar approach was developed, applying diffusion models to di-jet generation for weakly supervised searches using low level jet constituents~\cite{Buhmann:2023acn}.
However, in addition to low level generation, our work studies the application to high level variables as well as the first application of partial diffusion in high energy physics.
Other previous work applying diffusion models for anomaly detection has seen a diffusion model trained as an unsupervised tagger for identifying anomalous jets~\cite{Mikuni:2023tok}.

    \section{Dataset}
    The LHCO R\&D dataset~\cite{LHCOlympics} has become a standard candle of performance for resonant anomaly detection approaches.
It comprises background data from di-jet events produced in QCD interactions, and signal data from the all-hadronic decay of a massive resonant BSM particle into two lighter BSM particles.
The masses of the three BSM particles in the signal events are $m_{W^\prime} = 3.5$~TeV, $m_{X} = 500$~GeV, and $m_{Y} = 100$~GeV.

Signal and background events are generated with \texttt{Pythia}~8.219~\cite{Sjostrand:2014zea} with detector simulation performed with \texttt{Delphes}~3.4.1~\cite{deFavereau:2013fsa}.
Large radius jets, with a radius parameter $R=1.0$, are reconstructed using the anti-$k_t$ clustering algorithm~\cite{AntiKt} using the \texttt{FastJet}~\cite{Cacciari:2011ma} package.
There are in total 1~million background and 100,000 signal events.

The LHCO Black Box 2 sample, which also comprises 1~million events, is used as an alternative background sample.
The generation chain follows the same prescription, except \texttt{Herwig++}~\cite{Bahr:2008pv} is used to generate the events, with a slightly modified detector parametrisation.

Events are required to have exactly two jets, where at least one jet has $p_\mathrm{T}^{J} > 1.2$~TeV, and are ordered by decreasing mass.
% In order to remove the turn on in the \mjj distribution arising from the jet selections, we only consider events with $\mjj > 2.7$~TeV.
% To construct the training datasets we use varying amounts of signal events mixed in with the QCD dijet data.
In order to compare \drapes to other methods, we use the same high level input features as in Refs.~\cite{curtains,LaCathode,feta,interplay},
namely
\begin{equation*}
    m_{JJ},\, m_{J_1},\, \Delta m_{J} = m_{J_1} - m_{J_2},\, \tau_{21}^{J_1},\, \tau_{21}^{J_2},\, \Delta R_{JJ}=\sqrt{\Delta\eta^2 + \Delta\phi^2}.
\end{equation*}
% Here $\tau_{21}$ is the N-subjettinness~\cite{nsubjettiness} ratio of a large radius jet beyond two-prong or one-prong like,
% and $\Delta R_{JJ}=\sqrt{\Delta\eta^2 + \Delta\phi^2}$ is the angular separation of the two jets.
Furthermore, with diffusion models it is possible to operate on input sets, such as the constituent particles within a jet, with permutation equivariance.
As such, the model can be trained without enforcing any order on the constituents within the jet.
Therefore, in addition to the default feature set, we also consider the leading 128 constituents in each jet.
They are represented by their transverse momentum as a fraction of the jet transverse momentum and their coordinates in $\eta$-$\phi$ space relative to the central axis of the jet ($p_{\mathrm{T}}/p_{\mathrm{T}}^J$, $\Delta\eta$, $\Delta\phi$).

The background data follow a falling spectrum in \mjj above 2.7~TeV, with signal events localised as a resonance in \mjj with a peak at 3.5~TeV (corresponding to $m_{W^\prime}$), and mostly contained within a signal window $\mjj\in[3300,3700)$,
which we define as our signal region~(SR).
To improve the statistical accuracy when evaluating model performance, additional background events falling within the SR are also available for the high level input features~\cite{david_shih_2021_5759087}.

    \section{Method}
    In \drapes, we train diffusion models to learn a conditional generative model of the data from side-bands~(SB) surrounding our signal region.
The model is conditioned on \mjj and, under the assumption that the side-bands contain only background data, can be assumed to be a generative model for the background process across the full \mjj range, including the SR.
The \drapes models are trained on either the standard high level variables or the jet constituents of the two jets in the event.

For a nominal choice of side-bands we consider the region $\mjj\in[2800,6000)$, excluding the signal region.
Separate \drapes models are trained for a wide range of injected signal levels.

\subsection{Diffusion model}

We follow the `EDM' noise scheduler and network pre-conditioning introduced in Ref.~\cite{Karras2022}.
%  and succesfully applied to high energy physics data in \droid~\cite{pcdroid}.
Here, the probability ordinary differential equation defining the reverse diffusion process is given by
\begin{equation*}
    \diff x_t = - t \nabla_x \log p(x; t) \diff t,
\end{equation*}
where $p(x; t)$ is the distribution over the data at time $t$, where $t = \sigma \in [0, 80)$ (where $\sigma$ is referred to as the noise strength).

\begin{figure}[htbp]
    \centering
    \includegraphics[width=0.53\textwidth]{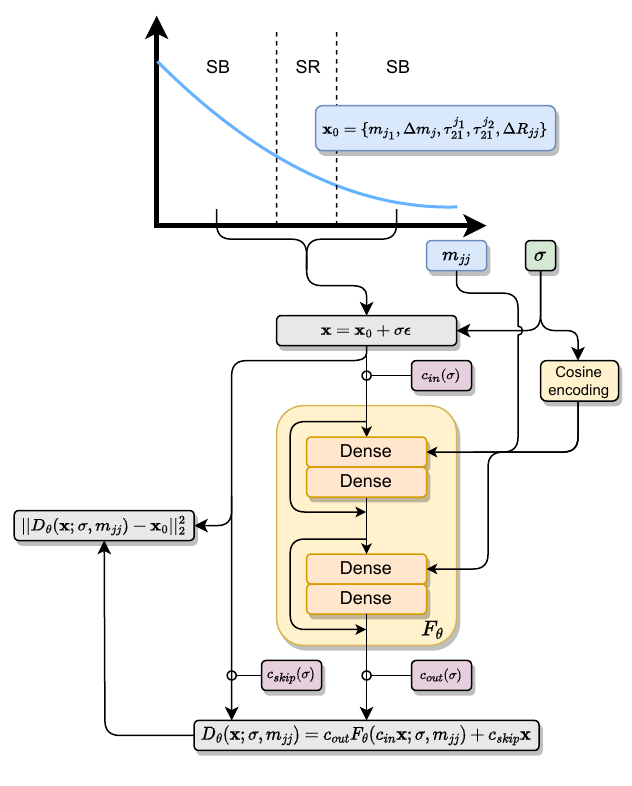}
    \caption{Schematic overview of the \drapes architecture during training when using high level variables.
    Data are drawn from the side-bands of the \mjj distribution, and their features are passed through the diffusion network $F_\theta$ after being perturbed with noise with strength $\sigma$.
    The network is conditioned on \mjj and $\sigma$.
    Skip connections modify the input and output of the network.
    % During generation new samples are either sampled purely from noise with $\sigma=80$ and iteratively denoised in a series of integration steps, or from events drawn from either the SBs, the SR, or an alternative reference sample, which have been perturbed with an initial noise strength $\sigma\in[0,80)$, and denoised in a series of integration steps.
    }
    \label{fig:arch}
\end{figure}

For high level input variables
% than considered in \droid, instead of using attention based transformer encoder layers,
we employ a simple ResNet~\cite{resnet} architecture comprising four hidden layers, each with 64 nodes, for the diffusion network.
The noise strength $\sigma$ is processed with a cosine embedding layer before being concatenated alongside \mjj as the conditional vector.
Skip and scaling connections, $c_\mathrm{in}(\sigma)$, $c_\mathrm{out}(\sigma)$ and $c_\mathrm{skip}(\sigma)$ are used in conjunction with the network, resulting in an output of
\begin{equation*}
    D_\theta(x;\sigma, \mjj) = c_\mathrm{out}(\sigma) F_\theta(c_\mathrm{in}(\sigma)x;\sigma, \mjj)+c_\mathrm{skip}(\sigma)x.
\end{equation*}
This results in a training objective of
\begin{equation*}
    \mathcal{L} = ||D_\theta(x;\sigma, \mjj)  - x_0||^2_2,
\end{equation*}
which is simply the distance between the network output and the true target, namely the original noised input data, $x_0$.
During training, we sample the noise rate following a shifted log-normal distribution
% \begin{equation*}
\mbox{$\log(\sigma)\sim\mathcal{N}\left(-1.2, 1.2\right)$}, and all inputs are scaled to have a mean of zero and unit variance.
The \drapes models are each trained for 700 epochs%
\footnote{Convergence is achieved after far fewer epochs, within an hour of training.}
with a learning rate rate of $10^{-4}$, with a linear warmup from zero over the first 50,000 training steps.
A schematic overview of the network architecture during training for the case of high level features is shown in \cref{fig:arch}.

% \end{equation*}

\subsection{Background sample generation}

Considering a single \drapes model, there are several options which can be used to generate reference background samples by performing the reverse diffusion process.

It is possible to use \drapes as a purely conditional generative model, generating new data for target values of \mjj.
To generate new samples noise is drawn from a multidimensional gaussian distribution, and the reverse diffusion process is performed starting at $\sigma=80$.
We refer to this method as {\boldmath\textbf{\drapesGen}}.
This generation method can be considered a diffusion analogue to the \CATHODE approach~\cite{cathode}, where samples are generated for target \mjj values with a normalizing flow starting from noise.

As with all diffusion models, the solver plays a key role in the fidelity and generation speed of samples.
Here we find performance is saturated with 50 integration steps, with no discernable difference between the Heun and Euler ordinary differential equation solvers.
Nonetheless, to ensure that the choice of solver plays a subleading role, we choose to use the Heun 2\textsuperscript{nd} solver with 200 integration steps at $\sigma=80$ when generating all samples.
%, as implemented with the \texttt{k-diffusion} library.%
% \footnote{\url{https://github.com/crowsonkb/k-diffusion/tree/v0.0.15}}

However, in addition to generating from pure noise, it is also possible to generate a reference background template by taking existing samples and using them as the inputs for partial diffusion.
In partial diffusion, the forward diffusion process is performed by adding noise to the samples corresponding to a target noise strength $\sigma^\prime$.
Then, the reverse diffusion process is performed for a target value \mjj, starting at $\sigma=\sigma^\prime$.
In this case, fewer passes are required to recover the output samples, as the diffusion process is starting partway along the trajectory.
The number of diffusion steps $N$ and $\sigma^\prime$ are related%
\footnote{As $N$ is required to be an integer, we instead use the inverse of the equation to determine the values of $\sigma^\prime$ which can be used.} %
by
\begin{equation}
    N = N_\text{max}\cdot{\sigma^\prime}^\frac{1}{\rho} - \frac{\sigma_\text{max}^\frac{1}{\rho}}{\sigma_\text{\vphantom{max}min}^\frac{1}{\rho} - \sigma_\text{\vphantom{min}max}^\frac{1}{\rho}}
\end{equation}
% \begin{equation*}
%     \sigma^\prime = \left(\sigma_\textrm{max}^\frac{1}{\rho} + \frac{N_\text{max}-N}{N_{\text{max}}-1}\left(\sigma_\textrm{min}^\frac{1}{\rho} - \sigma_\textrm{max}^\frac{1}{\rho}\right)\right)^\rho,
% \end{equation*}
where $N_\text{max}=200$ and $\sigma_{max}=\sigma=80$ is the total number of reverse diffusion steps and corresponding noise strength during training. Here $\sigma_\text{min}=10^{-5}$ and $\rho=7$ are hyperparameters of the diffusion model.
% In the partial diffusion case there are three approaches we consider.

We consider three different approaches for partial diffusion, which differ based on the source of the initial data.
\begin{description}
    % \item[\drapesGen] In the simplest case, new samples are generated for target values of \mjj by drawing $x$ from random noise and iteratively performing the reverse diffusion process starting at $\sigma=80$.
    % This case most closely represents the \CATHODE approach using normalizing flows.
    \item[\drapesSR] Here, partial diffusion is applied to the data in the signal region with a target noise strength $\sigma^\prime$.
    They are subsequently denoised using the reverse diffusion process using the original \mjj values.
    \item[\drapesSB] Here, data are drawn from the side-bands and partial diffusion is applied for a target \mjj value which lies in the SR, rather than their initial \mjj value.
    This case most closely resembles the \CURTAINs method which uses normalizing flows~\cite{curtains,curtainsf4f} to transform sideband data to target values of \mjj within the signal region.
    \item[\drapesMC] Here partial diffusion is applied to data in the signal region which are drawn from another reference sample, for example Monte Carlo simulation.
    Here we use the alternative background sample from the LHC Olympics.
    This case most closely resembles the \FETA approach~\cite{feta} which uses normalizing flows to transform signal region data from another sample to the signal region of the nominal sample.
\end{description}
{\flushleft
For the case of \drapesGen and \drapesSB, we sample target values of \mjj from a four-parameter polynomial
\begin{equation}
    f(z) = p_1 \left(1 - z \right)^{p_2} z^{p_3 + p_4 \log z},
\end{equation}
where $z=\mjj/\sqrt{s}$.
The parameters $p_i$ are extracted by performing a fit to the side-band data.
}

In all cases it is possible to oversample the reference background sample by generating more data than are present in the SR.
For \drapesSR and \drapesMC this is done by sampling different noise in the forward diffusion process multiple times for the same events.
For \drapesGen and \drapesSB, both the noise in the forward diffusion process and target \mjj values are sampled independently multiple times for each event.

\subsection{Drapes for jet constituents}

\begin{figure}[htbp]
    \centering
    \includegraphics[width=0.48\textwidth]{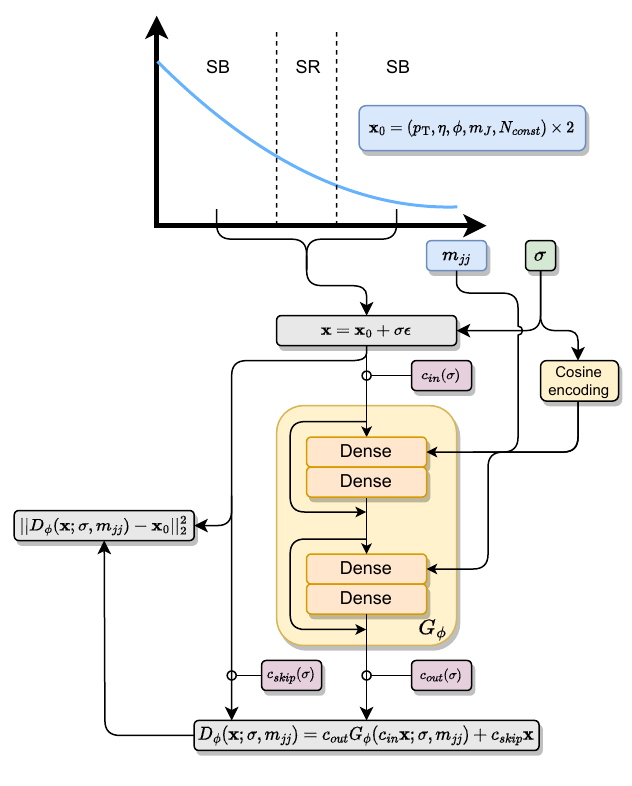}
    \includegraphics[width=0.48\textwidth]{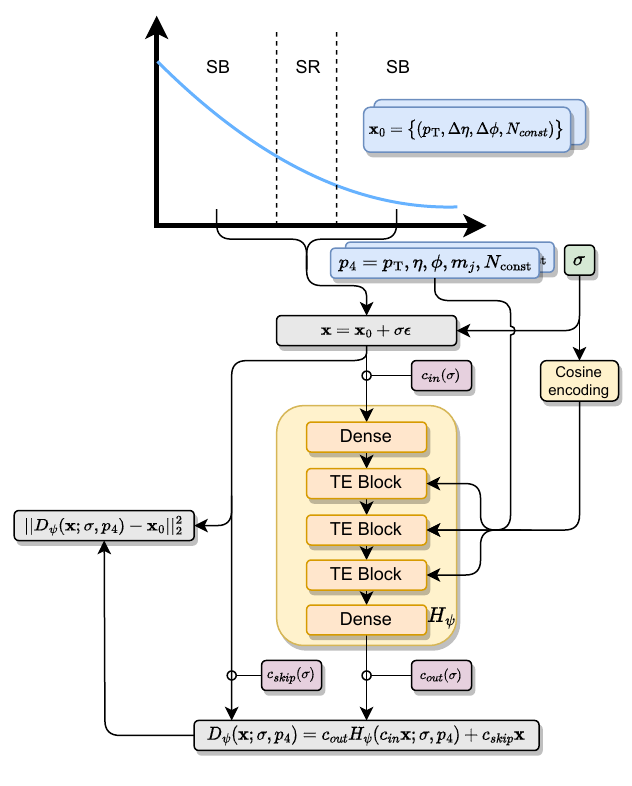}
    \caption{Schematic overview of the \drapes architecture during training when applied to jet constituents.
    The approach comprises two components, the di-jet kinematic model (left) and the per-jet model (right).
    Data are drawn from the side-bands of the \mjj distribution.
    To learn the di-jet kinematics as a function of \mjj, the four-momentum and number of jet constituents for each jet $p_4$ are passed through the kinematic diffusion network $G_\phi$ after being perturbed with noise with strength $\sigma$.
    The network is conditioned on \mjj and $\sigma$.
    To learn the jet constituents as a function of $p_4$, the constituents of each jet are passed through the jet diffusion network $H_\psi$ after being perturbed with noise with strength $\sigma$.
    The network is conditioned on $p_4$ and $\sigma$.
    In both networks skip connections modify the input and output of the network, and partial diffusion can also be employed.
    % New samples are either sampled purely from noise with $\sigma=80$ and iteratively denoised in a series of integration steps, or from events drawn from either the SBs, the SR, or an alternative reference sample, which have been perturbed with an initial noise strength $\sigma\in[0,80)$, and denoised in a series of integration steps.
    }
    \label{fig:archLL}
\end{figure}

Due to the flexibility of diffusion models in comparison to normalizing flows, \drapes can easily be adapted to any type of input.
To apply \drapes to low level jet constituents, we generate the constituents of each jet in isolation using the same diffusion model.
The diffusion model builds upon the architecture and prescription used in Ref.~\cite{pcdroid} with the denoising architecture built from a series of transformer encoder~(TE) blocks~\cite{vaswani2023attention}, without a further optimisation of the hyperparameters.
The jet four-momentum and number of jet constituents ($p_4 = (\pt, \eta, \phi, m_j, N_{const})$)%
\footnote{For notational convenience we include $N_{const}$ in the definition of the four-vector.}
 are provided as conditional information.
However, in order to obtain the conditional information $p_4$ for each jet, we require a second diffusion model conditional on \mjj.
This network is identical to the \drapes architecture used for the high level features.
The two networks comprising the low level \drapes model are shown in \cref{fig:archLL}.

In background QCD events we do not expect significant correlations between the substructure of the two jets, which are produced as either two gluons, two quarks, or one of each, predominantly via t-channel.%
\footnote{At the LHC, both pairs of gluons and pairs of quarks are more likely to be produced than a gluon-quark pair.
However, due to the similarity between quarks and gluons as well as contributions from additional radiation and next to leading order production mechanisms, we expect this to be a subleading effect, especially in comparison to any potential resonant signal.}
However, for any resonant signal process the two daughter particles will be related and their substructure will be strongly correlated.
%By applying \drapes to each jet in isolation we do not preserve these correlations and could introduce more sensitivity even in the presence of signal events in the sideband data.
By applying \drapes to each jet in isolation, in the case where there are signal events in the sidebands the substructure of signal jets could be learned, but the correlations between the substructure of the two jets would not be.
Alternatively, if \drapes were applied to the constituents in both jets simultaneously, the correlations between the substructure of signal jets would be learned.
As a result, when generating reference templates for the signal region, considering each jet independently should have fewer signal like events, where both jets have signal like substructure, resulting in better sensitivity than when the two jets are not independent.

Even in the case where there are no signal events present in the sidebands, applying \drapes to each jet in isolation should still be beneficial for partial diffusion.
After the addition of noise and applying independent denoising processes to each jet, the correlations between the substructure of two signal jets in an event would be decreased.
% Breaking these correlations in signal events could also be exploited  using partial diffusion.

% Moving from normalizing flows to diffusion models opens up the possibility to use generative models on low level inputs rather than just high level variables.

% Unlike normalizing flow based approaches, which are restricted to low dimensional structured inputs, diffusion models can utilise more complex and expressive network architectures and remain invariant to symmetries in input data.
% Diffusion models have been successfully used to generate the constituents of jets and calorimeter shower, achieving state of the art performance.

% As a result, not only can \drapes be applied to high level variables previously used by normalizing flow based approaches, it can also be used in conjunction with transformer based architectures applied to jet constituents.

% In comparison to normalizing flows, which are well suited to low dimensional structured inputs, but constrained in choice of network architectures, diffusion models have more freedom in the choice of architecture

\subsection{Weakly-supervised classifiers}

To test the performance of the various approaches, we train \CWoLa classifiers to separate the signal region data from the generated templates.
The SR data can either be purely background events, or contain a controlled number of signal data, which can also contaminate the side-bands.
As the classifiers themselves are not the focus on this work, we use the same hyperparameters and architecture used in Refs.~\cite{curtains,curtainsf4f} without further optimisation for the high level features, despite promising performance improvements observed with decision trees~\cite{Finke:2023ltw,Freytsis:2023cjr} or for mitigating sculpting after applying cuts~\cite{LaCathode}.
This enables easier direct comparison with \FfF.
A five-fold cross validation approach is employed to train all classifiers, including the fully supervised configuration.

Following the \CWoLa classifiers for the high level features, we keep the architecture small also for the low level approach with \drapes applied to generate jet constituents.
We use a classifier which is permutation invariant to both the jet constituents, but also the two jets.
To achieve this we use transformer encoders operating on the jet constituents with self-attention~\cite{vaswani2023attention} layers.
The output is then pooled with a learnable class token cross-attention~\cite{touvron2021going} layer, weighting the contribution from each constituent; this ensures permutation invariance over the jet constituents.
The same transformer is applied to the constituents of both jets, with the outputs combined in a summation in order to ensuring permutation invariance between the two jets.
The resulting vector is subsequently processed with a small multi-layer perceptron.

\begin{figure}[htbp]
    \centering
    \includegraphics[width=0.75\textwidth]{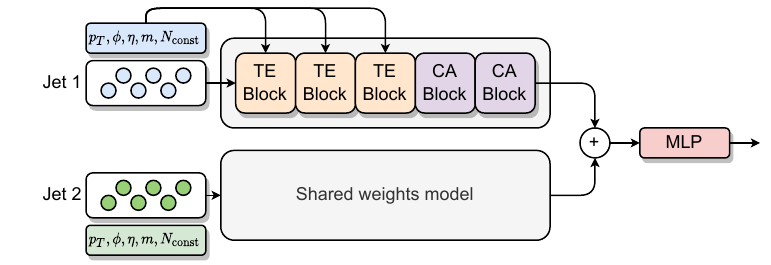}
    \caption{Schematic overview of the weakly supervised classifier for the jet constituents.
    The same transformer network is applied to the constituents of both jets, with the outputs combined in a summation in order to ensuring permutation invariance between the two jets.
    }
    \label{fig:archLLclass}
\end{figure}

\subsection{Comparison with prior work}

As previously discussed, various modes of sample generation with \drapes are analogous to existing approaches but using diffusion models rather than normalizing flows.
However, during the development of this work Ref.~\cite{Buhmann:2023acn} applied flow matching and diffusion models to the task of dijet generation in a signal region.
This approach is equivalent to \drapesGen applied to the low level jet constituents, however, this work differentiates itself in several key regards with respect to Ref.~\cite{Buhmann:2023acn}.

With \drapes, we introduce and study the use of partial diffusion as a new means of generating samples, rather than generating only from noise as with \drapesGen.
Furthermore, we demonstrate that diffusion models can outperform normalizing flow based approaches on the same features, and that there are advantages beyond the more flexible choice of architecture that enables jet constituent generation.
We compare the performance of \drapes applied to high and low level variables across a wide range of injected signal events, and make comparisons for the case of partial diffusion for both sets of inputs.
% Finally, the classifier employed for the jet constituent based models is completely permutation invariant, also with respect to the order of the two jets.
% This is thanks to the summation rather than concatenation operation.

    \section{Results}
    In order to measure how well \drapes generates reference background samples,
we focus on a single signal region defined as $\mjj \in [3300,3700)$ with various levels of signal contamination.
% In all cases we train a \CWoLa classifier to separate the reference from the true data in the SR.
% A five-fold cross validation approach is employed to train all classifiers following the approach in Ref.~\cite{curtainsf4f}.
The main measure of performance of the subsequent classifier is the significance improvement, given by $\rfrac{\epsilon_S}{\sqrt{\epsilon_B}}$, where $\epsilon_S$ and $\epsilon_B$ are the signal and background efficiencies respectively.
Similar to $\epsilon_B$ in a receiver operator characteristic~(ROC) curve, we calculate the significance improvement across the full range of background rejection values ($\rfrac{1}{\epsilon_B}$).
% The significance improvement measures the expected improvement in statistical sensitivity in a bump hunt over not applying any cut on the data.

We compare the performance of \drapes in several scenarios to a \FfF model trained using the full side-band width, which was shown in Ref.~\cite{curtainsf4f} to deliver state of the art performance.
For reference we also show the performance of a supervised classifier trained to separate the signal process from the QCD background on an independent set of data, and an idealised \CWoLa classifier.
In the idealised classifier, simulated QCD background events are used as the reference background sample, with oversampling equal to the amount used in \drapes and \FfF (Over-Idealised).

\subsection{Simple diffusion}

\begin{figure}[hbpt]
    \centering
    \begin{subfigure}{0.48\textwidth}
        \centering
        \includegraphics[width=\textwidth]{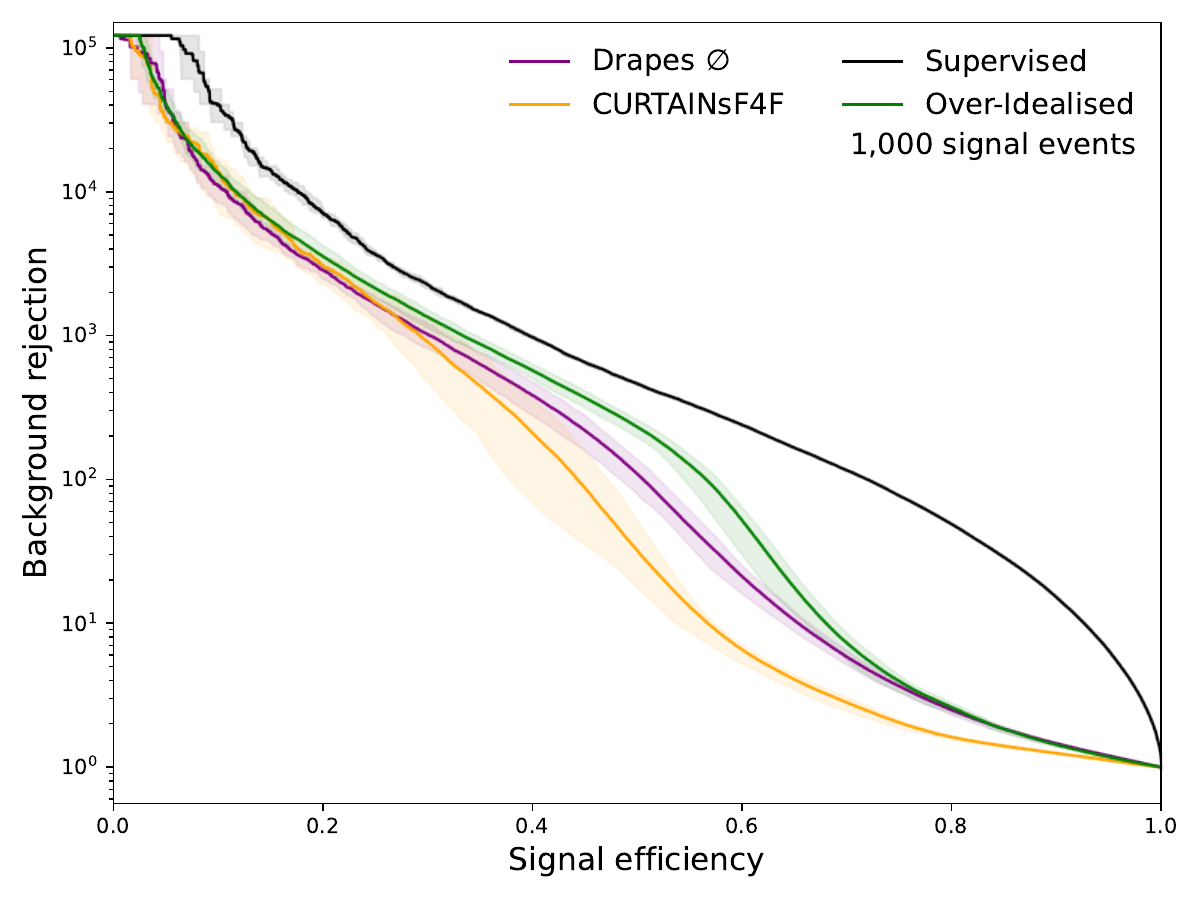}
    \end{subfigure}
    \hfill
    \begin{subfigure}{0.48\textwidth}
        \centering
        \includegraphics[width=\textwidth]{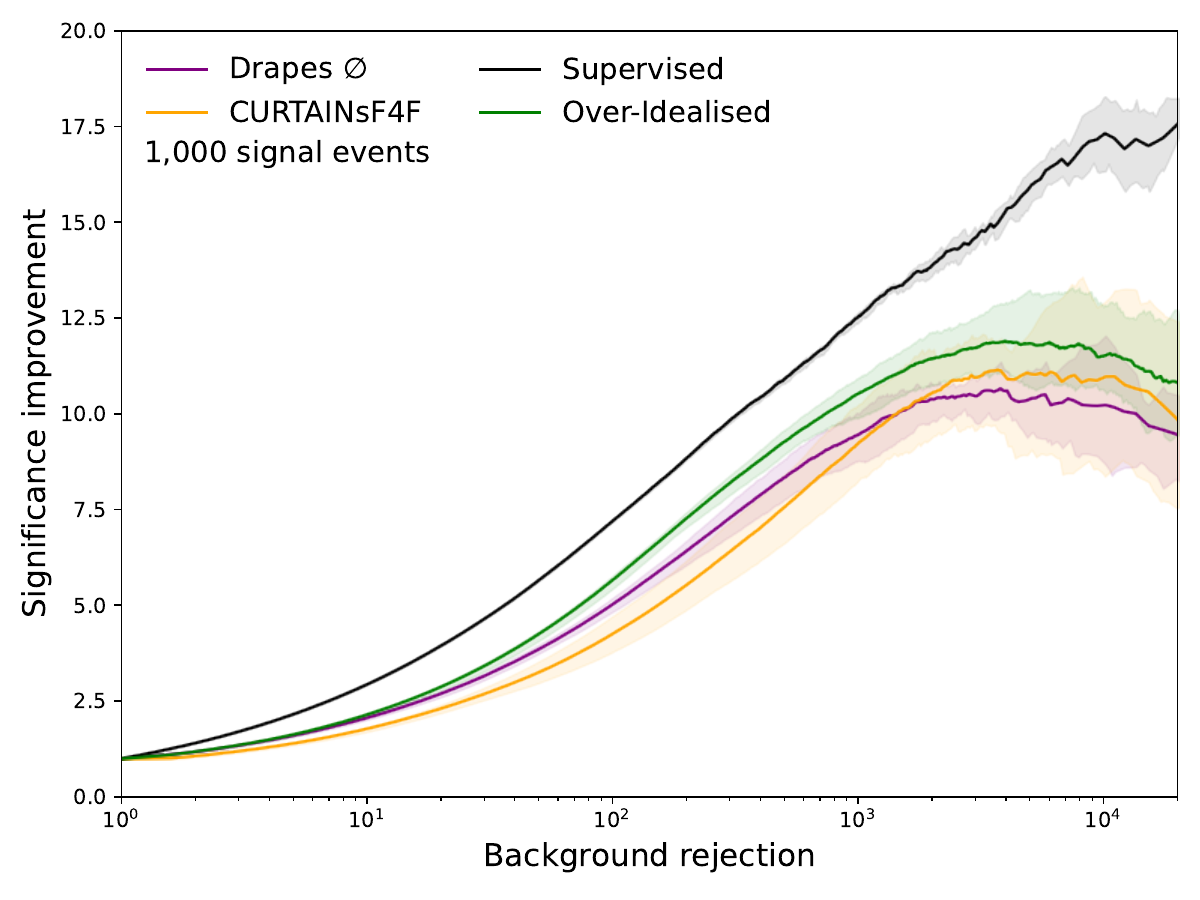}
    \end{subfigure}\\
    \begin{subfigure}{0.48\textwidth}
        \centering
        \includegraphics[width=\textwidth]{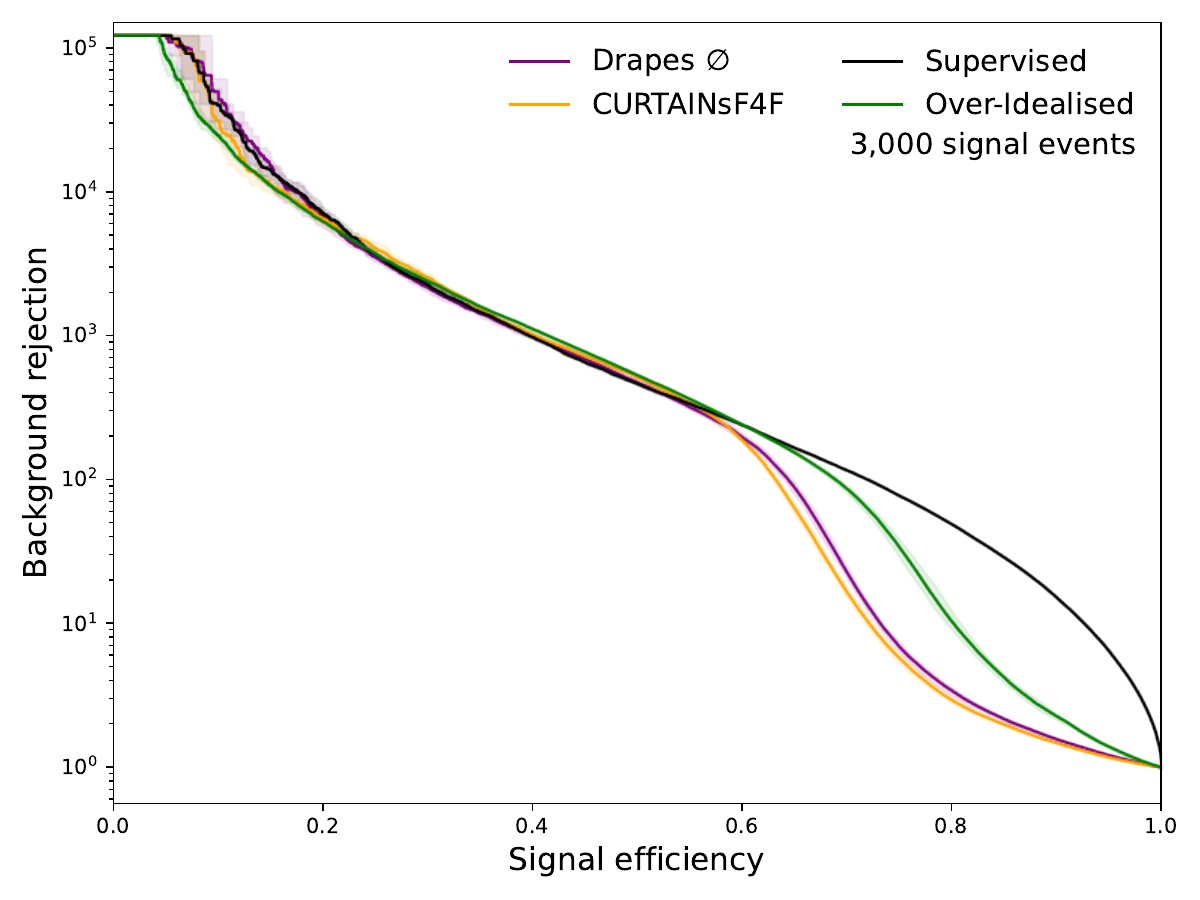}
    \end{subfigure}
    \hfill
    \begin{subfigure}{0.48\textwidth}
        \centering
        \includegraphics[width=\textwidth]{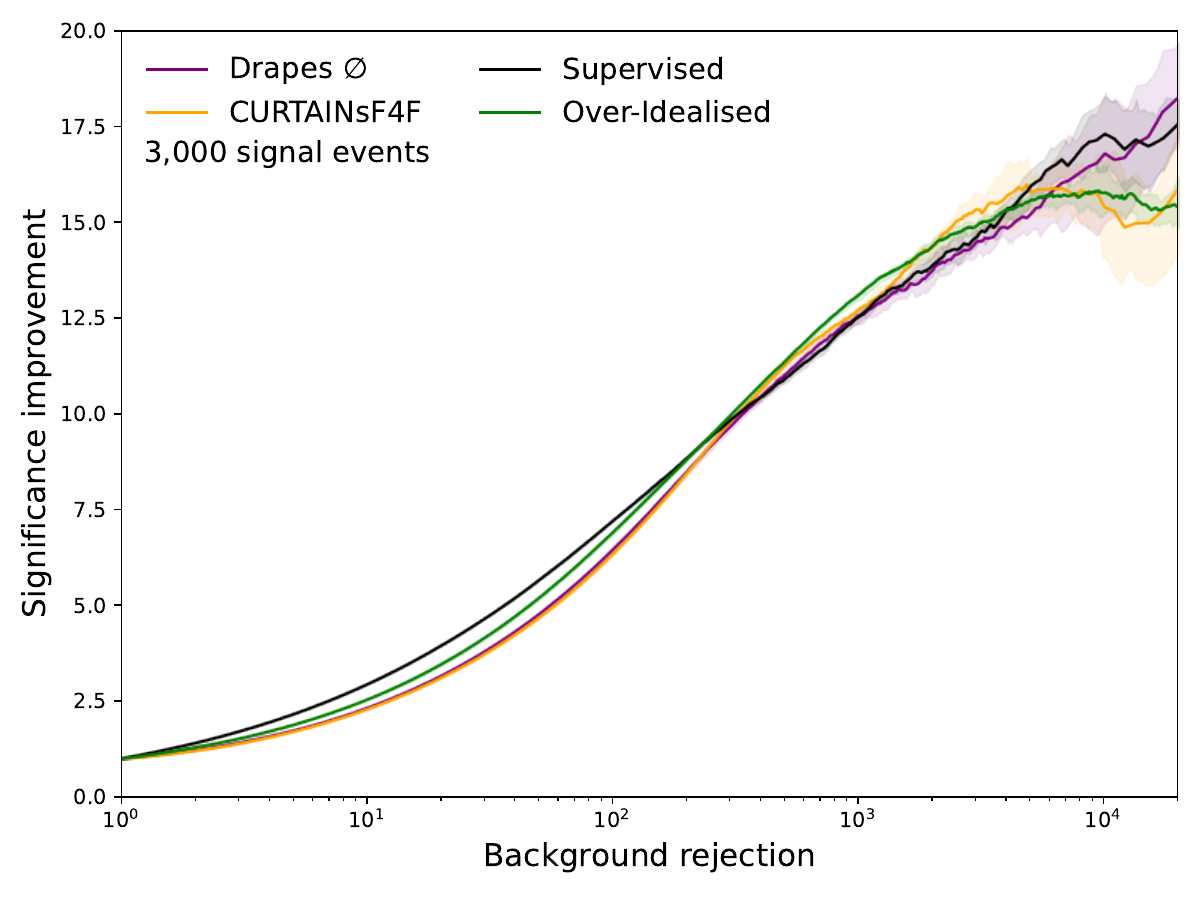}
    \end{subfigure}
    \caption{Background rejection as a function of signal efficiency (left) and significance improvement as a function of background rejection (right) for \drapesGen~(purple), \FfF~(orange), Supervised~(black), and Over-Idealised~(green).
    All methods are trained on the sample with 1,000 (top) and 3,000 (bottom) injected signal events, and a signal region $3300\leq\mjj<3700$~GeV.
    The lines show the mean value of fifty independent classifiers, with the shaded band representing a 68\% uncertainty.
    % Supervised and idealised classifiers are trained for reference.
    }
    \label{fig:sic_roc_simple}
\end{figure}

As it is the simplest approach, we first look at the performance of \drapesGen.
In \cref{fig:sic_roc_simple} we look at the performance of the \CWoLa classifiers for two levels of signal injection.
Where 1,000 signal events are present in the data, we see that  \drapesGen achieves state-of-the-art performance as a purely generative model.
Here, \drapesGen has higher significance improvement values than the Over-Idealised classifier only due to the different profile of the ROC curve.
The Over-Idealised classifier has greater separation power, with an AUC of 0.89 in comparison to 0.85 with \drapesGen.
Similarly we see that \drapesGen reaches state-of-the-art performance where 3,000 signal events have been injected.
However, we note here that both \drapesGen and \FfF reach saturation in significance improvement at high levels of background rejection, achieving the same level of performance of the supervised classifier.

\begin{figure}[hbpt]
    \centering
    \begin{subfigure}{0.49\textwidth}
        \centering
        \includegraphics[width=\textwidth]{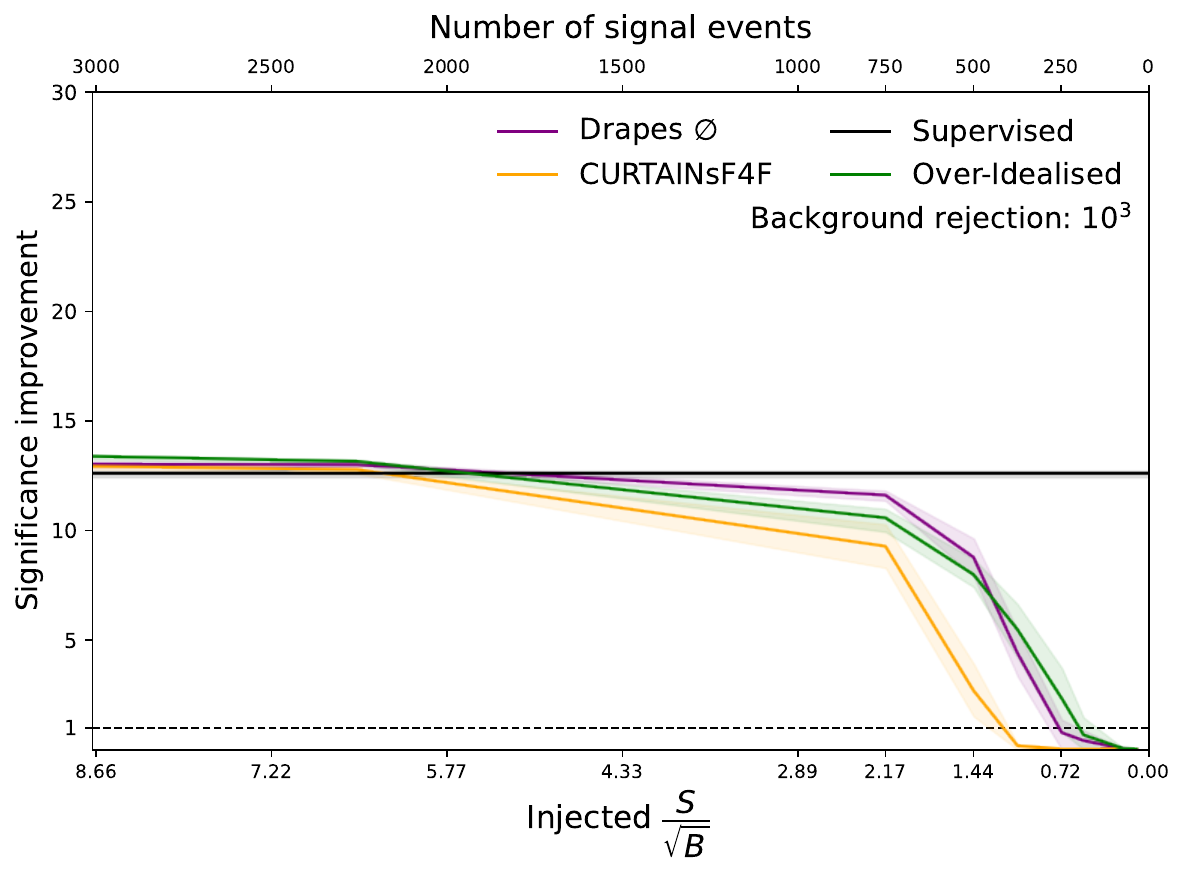}
    \end{subfigure}
    \begin{subfigure}{0.49\textwidth}
        \centering
        \includegraphics[width=\textwidth]{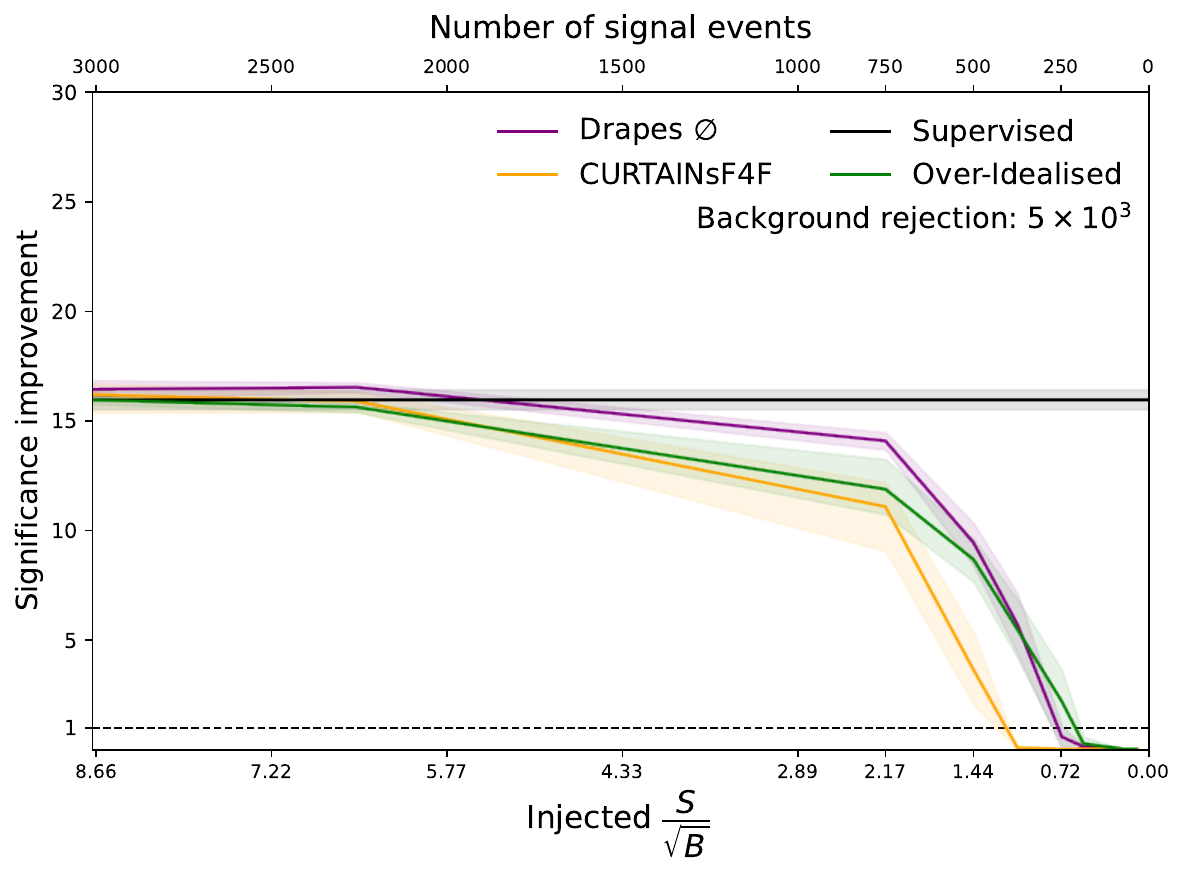}
    \end{subfigure}
    \caption{Significance improvement at fixed background rejection rates of $10^3$ (left) and \mbox{$5 \times 10^3$} (right) as a function of the number of signal events in the signal region, \mbox{$3300\leq\mjj<3700$~GeV}, for \drapesGen~(purple), \FfF~(orange), 
    %\drapesSR~(red), \drapesSB~(blue), \drapesMC~(magenta),
    Supervised~(black), and Over-Idealised~(green).
    The injected significance measured from the number of signal and background events in the signal region before applying a cut $\frac{S}{\sqrt{B}}$ is also shown.
    The lines show the mean value of fifty independent classifiers, with the shaded band representing a 68\% uncertainty.
    }
    \label{fig:sic_vs_sig_gen}
\end{figure}

In \cref{fig:sic_vs_sig_gen} we look at the performance of \drapesGen as the number of signal events present in the data changes.
In comparison to \FfF we see that \drapesGen is able to remain more sensitive at lower $\rfrac{S}{B}$.
However, \drapesGen drops off in performance below injected significances of 1$\sigma$ at background rejections of $10^3$, where fewer than 350 signal events are present in the signal region.
This is just before the Over-Idealised classifier indicates the limit of the \CWoLa weakly supervised approach.
The same trend is observed for a background rejection factor of $5\times10^3$. %, where some significance improvement is still observed at $\rfrac{S}{B} = 0.2\%$.

\subsection{Partial diffusion}

When using \drapes with performing partial diffusion, it is critical that the generated reference distributions still match the true background distribution.
For example, when starting from side-band data with a value of $\sigma^\prime$ too close to zero, the overall distributions will not change and a \CWoLa classifier will identify the differences between the two distributions arising from the correlations to \mjj instead of differences between the signal present in one class, and the approximation of the background data in the other.
Similarly, when starting from the signal region data themselves, if $\sigma^\prime$ is too small, the samples will be statistically identical and the classifier will not be able to separate them, leading to a classifier unable to separate signal events from the background.

We measure the performance of the partial diffusion generation as we change $\sigma^\prime$ by training \drapes on a sample with no signal present.
By generating background reference templates for a wide range of $\sigma^\prime$ we can then test how close they match the true background by training a classifier to separate them.
If the area under the ROC curve of the classifier is close to 0.5, then the reference samples are good approximations of the background.
However, if the values are too large, a \CWoLa classifier would instead focus on this mismodelling of the background.

\begin{figure}[hbpt]
    \centering
    \begin{subfigure}{0.5\textwidth}
        \centering
        \includegraphics[width=\textwidth]{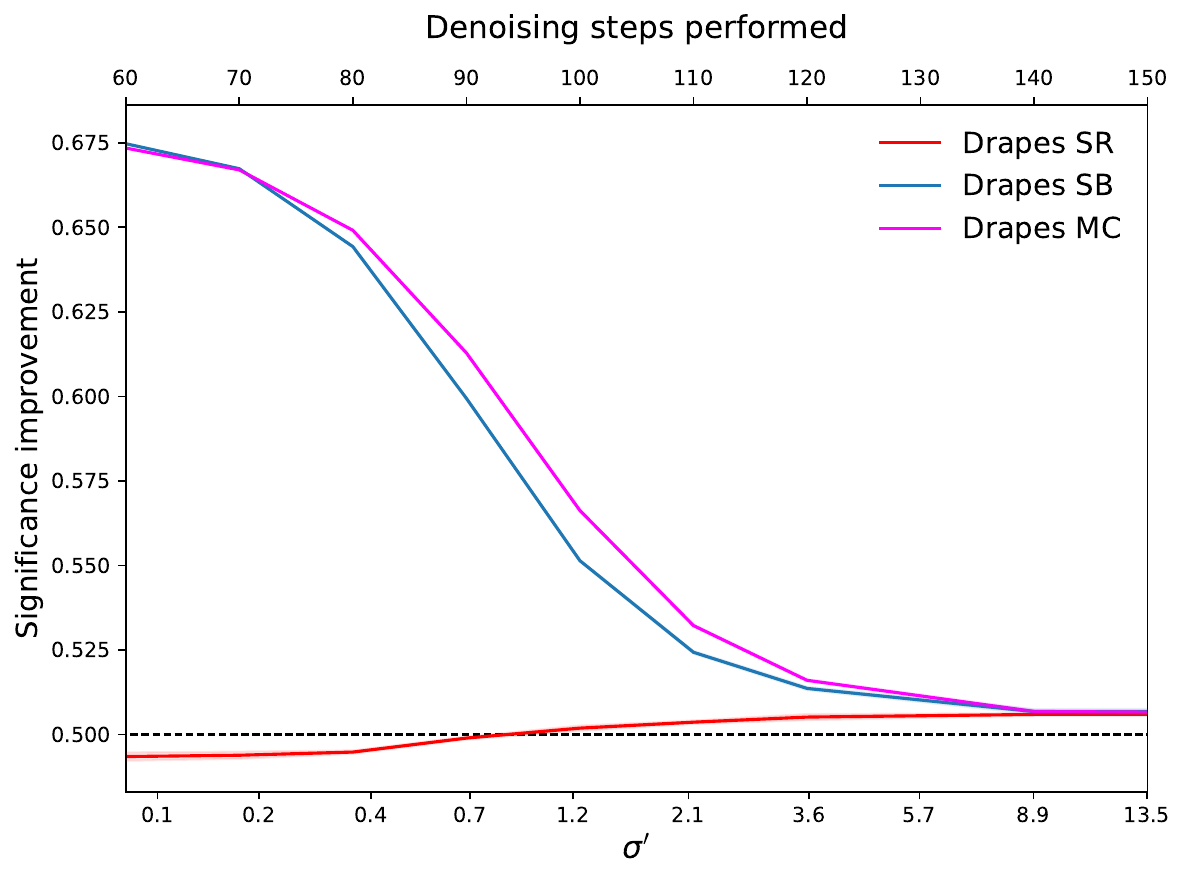}
    \end{subfigure}
    \caption{AUC values of classifiers trained to separate the \drapes reference samples from data in the case where no signal has been injected.
    The AUC is calculated as a function of the initial $\sigma^{\prime}$ used to generate samples using \drapesSR~(blue), \drapesMC~(magenta) and \drapesSB~(red).
    By construction, all methods converge to the same outputs as $\sigma^\prime$ increases, and at a value of $\sigma^\prime=80$ all methods are equivalent when using the same \mjj values.
    An AUC of 0.505 is observed at the convergence of the methods.
    The lines show the mean value of fifty independent classifiers, with the shaded band representing a 68\% uncertainty, which are negligible at this scale.
    }
    \label{fig:closure_aucs}
\end{figure}

In addition, we compare the performance of \drapesGen with partial diffusion, starting with data drawn from noise but with fewer denoising steps and the same values of $\sigma^\prime$.
The AUC scores for all four \drapes methods are compared in \cref{fig:closure_aucs}.
We can see that all three methods starting from reference data, \drapesSR, \drapesSB, and \drapesMC, result in reference samples much closer to the true background distribution than when starting from noise with \drapesGen.
This shows that there is a benefit for starting with some data rather than just pure noise, but only up until a point, as after some value of $\sigma^\prime$ the input data are almost indistinguishable from noise.
By construction, \drapesSR is closest to the true background distribution as they are used as the starting samples.
For \drapesSR and \drapesSB $\sigma^\prime$ values above 0.62 result in reference templates with AUC values less than 0.55, which are generally close enough to result in reasonable \CWoLa classifiers in the presence of signal.

\begin{figure}[hbpt]
    \centering
    \begin{subfigure}{0.48\textwidth}
        \centering
        \includegraphics[width=\textwidth]{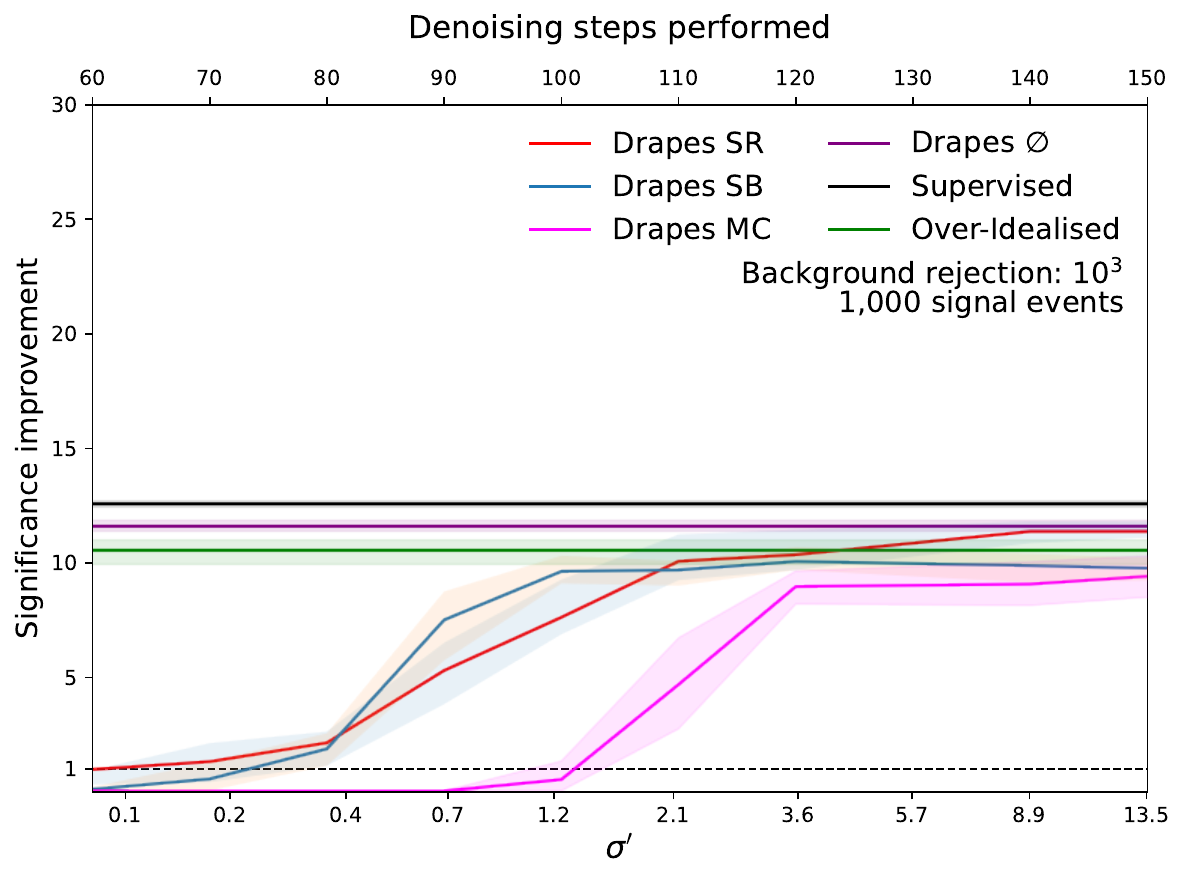}
    \end{subfigure}
    \hfill
    \begin{subfigure}{0.48\textwidth}
        \centering
        \includegraphics[width=\textwidth]{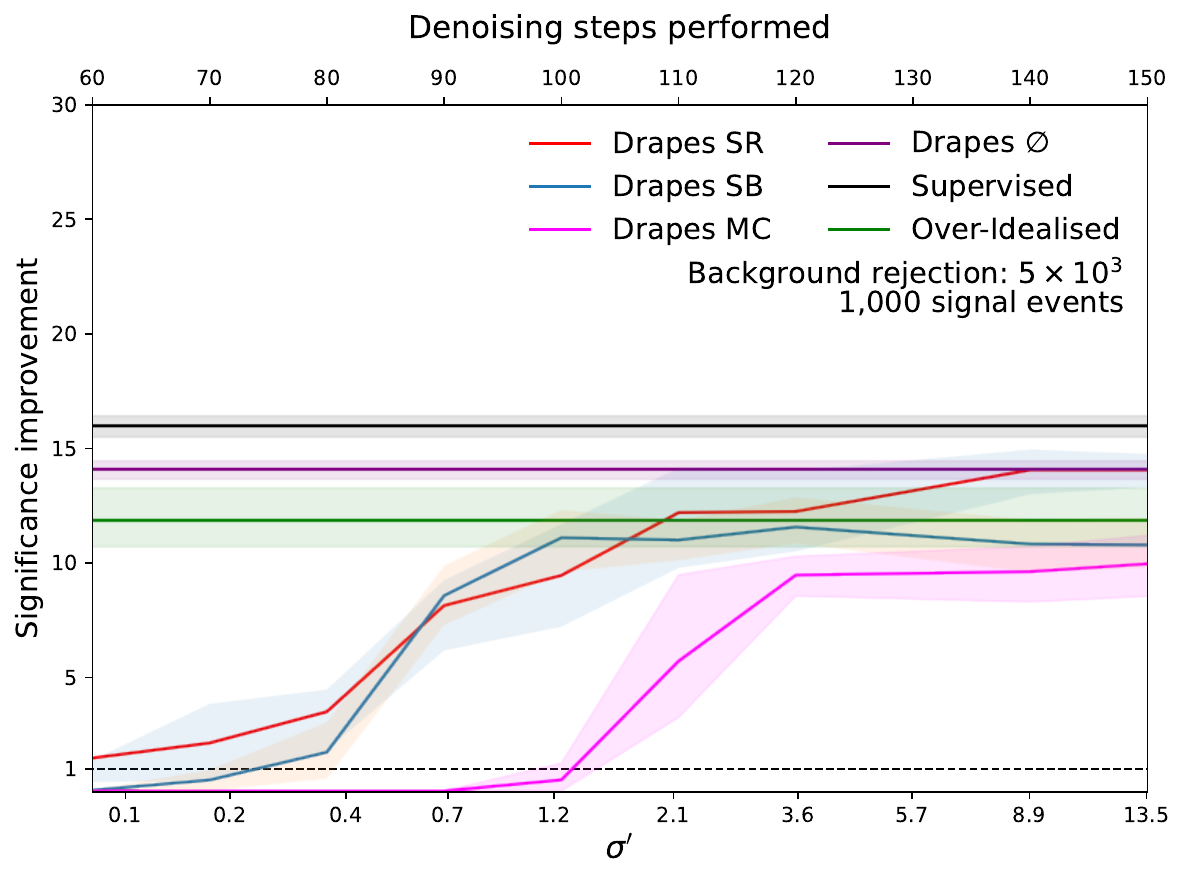}
    \end{subfigure}\\
    \begin{subfigure}{0.48\textwidth}
        \centering
        \includegraphics[width=\textwidth]{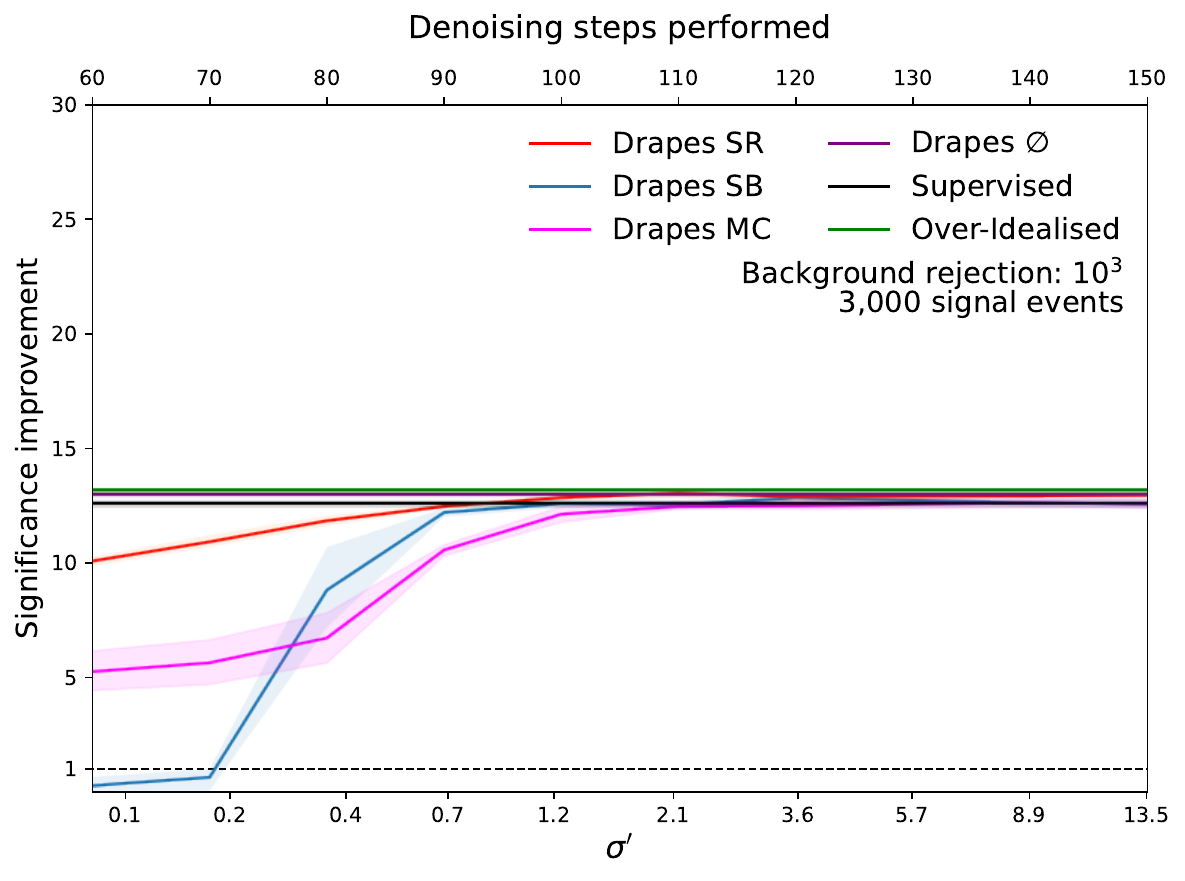}
    \end{subfigure}
    \hfill
    \begin{subfigure}{0.48\textwidth}
        \centering
        \includegraphics[width=\textwidth]{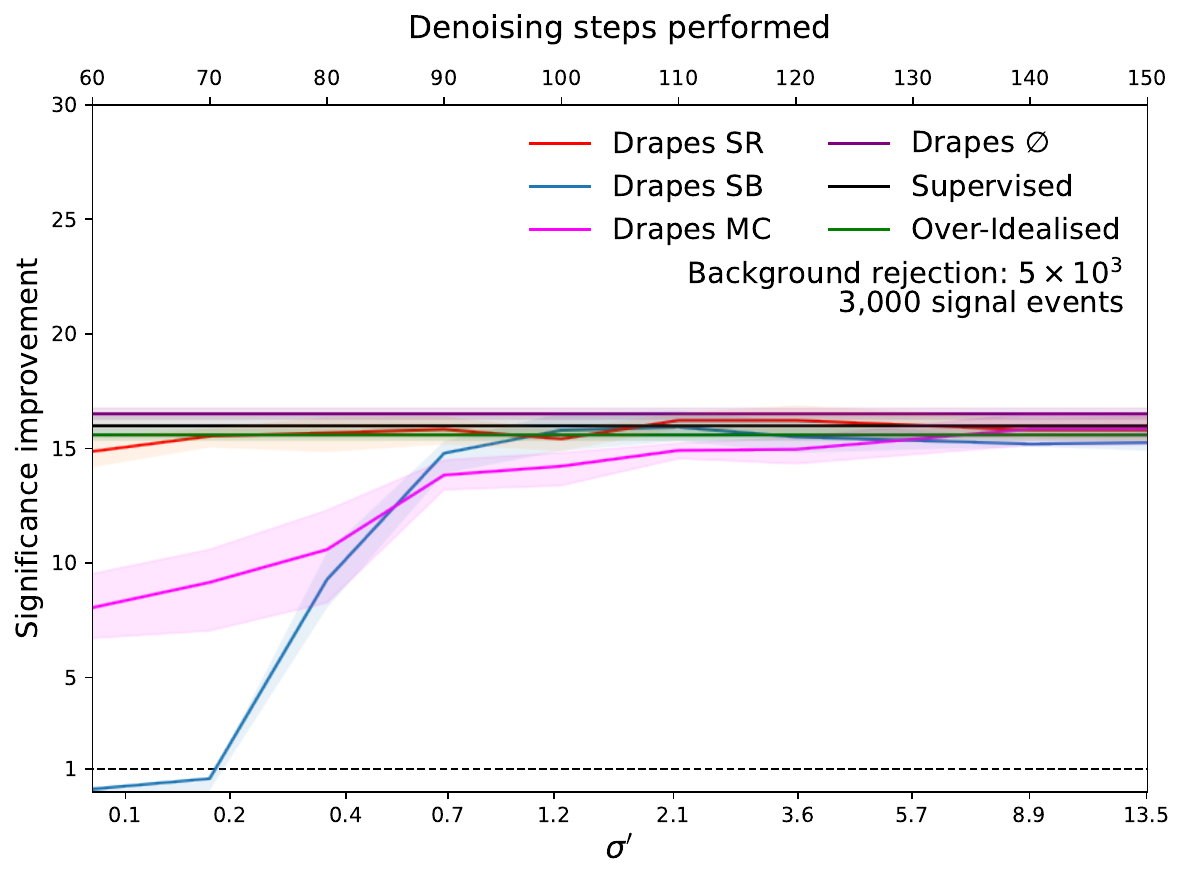}
    \end{subfigure}
    \caption{Significance improvement as a function of noise rate for \drapesSR~(red), \drapesSB~(blue), and \drapesMC~(magenta).
    % \drapesGen~(purple) is shown for reference, with initial samples always drawn from noise.
    All generation methods use the same diffusion model trained on the sample with 1,000 (top) and 3,000 (bottom) injected signal events, and a signal region $3300\leq\mjj<3700$~GeV.
    % A noise rate of 0 represent the input data without any noise perturbation or diffusion steps, and a rate of 1 corresponds to the case where the inputs are completely noise.
    % In the case of \drapesSB target \mjj values for each sample are drawn from a polynomial fit, whereas in \drapesSR the original value for \mjj is used.
    Reference values for \drapesGen~(purple), Supervised~(black) and Over-Idealised~(green) are shown in addition.
    The lines show the mean value of fifty independent classifiers, with the shaded band representing a 68\% uncertainty.
    }
    \label{fig:sic_noise}
\end{figure}

In \cref{fig:sic_noise} the performance of the partial diffusion approaches are compared as we vary $\sigma^\prime$ in the case of 1,000 and 3,000 injected signal events.
As expected, in the case of fewer signal events the overall performance is more sensitive to the modelling of the background reference template.
For 1,000 signal events we see that both \drapesSR and \drapesSB maintain competitive performance for values of $\sigma^\prime$ down to 0.6, where the AUC scores are also below 0.55, however for lower values the performance degrades.
Interestingly, \drapesMC has lower overall performance and requires more denoising steps to reach competitive performance.
\drapesSR and \drapesSB have similar performance for 1,000 injected signal events, though \drapesSR can maintain performance with far fewer steps in the presence of more signal in the SR.

In order not to optimise for $\sigma^\prime$ for all amounts of signal and subsequent studies we choose a single value for $\sigma^\prime$ using the AUC in the no signal case shown in \cref{fig:closure_aucs}.
Here we see that the agreement of the partial diffusion approaches with true background data saturates at around $\sigma^\prime=2.24$.
This also corresponds to where the performance has saturated for all three methods at the case of 1,000 signal events in \cref{fig:sic_noise}.
It should be noted that that this may not be the most optimal choice for all methods at low signal levels.
% Equally, where the focus is on faster generation, choosing a lower value would result in fewer network evaluations which, depending on the amount of signal present, may not result in a large change in overall sensitivity.

% For given choices of $\sigma^\prime$ we can observe how the partial diffusion approaches perform in comparison to \drapesGen for a range of signal ranges.

\begin{figure}[hbpt]
    \centering
    \begin{subfigure}{0.49\textwidth}
        \centering
        \includegraphics[width=\textwidth]{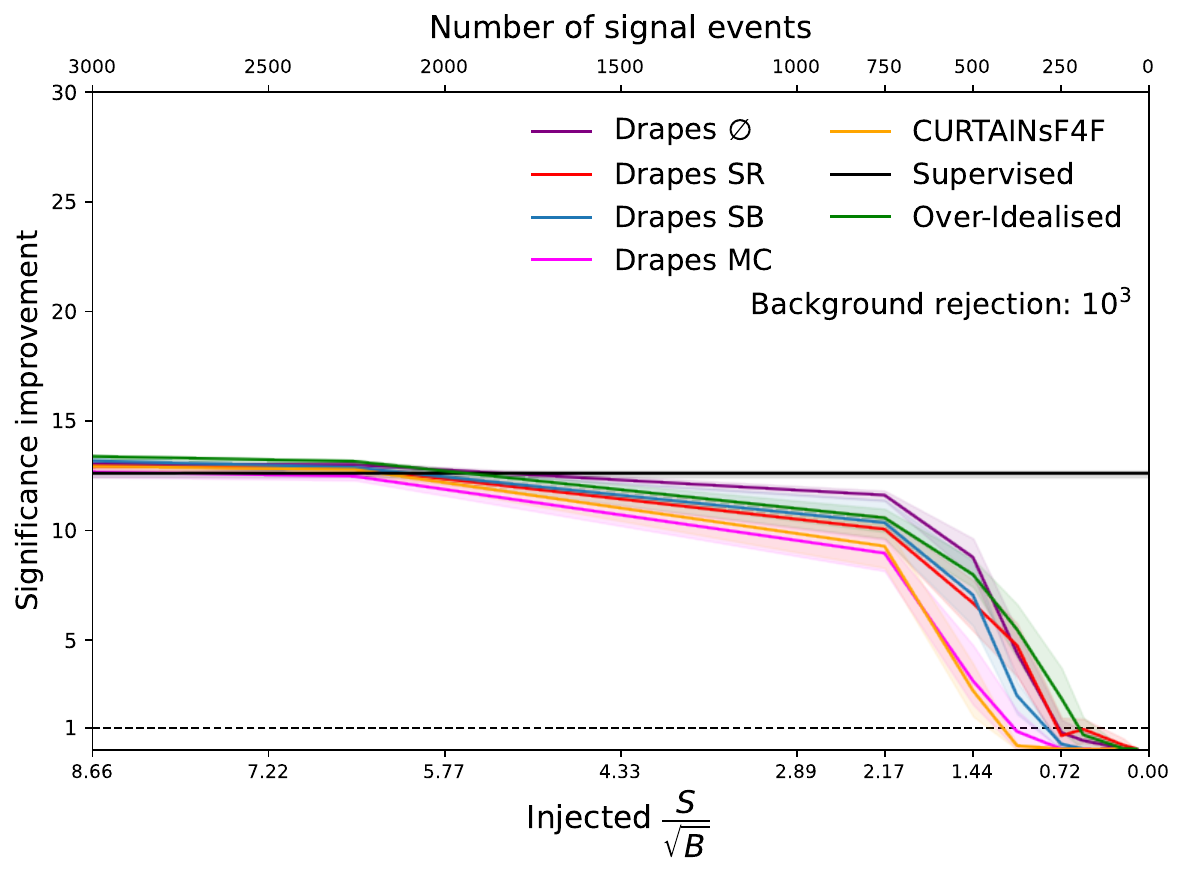}
    \end{subfigure}
    \begin{subfigure}{0.49\textwidth}
        \centering
        \includegraphics[width=\textwidth]{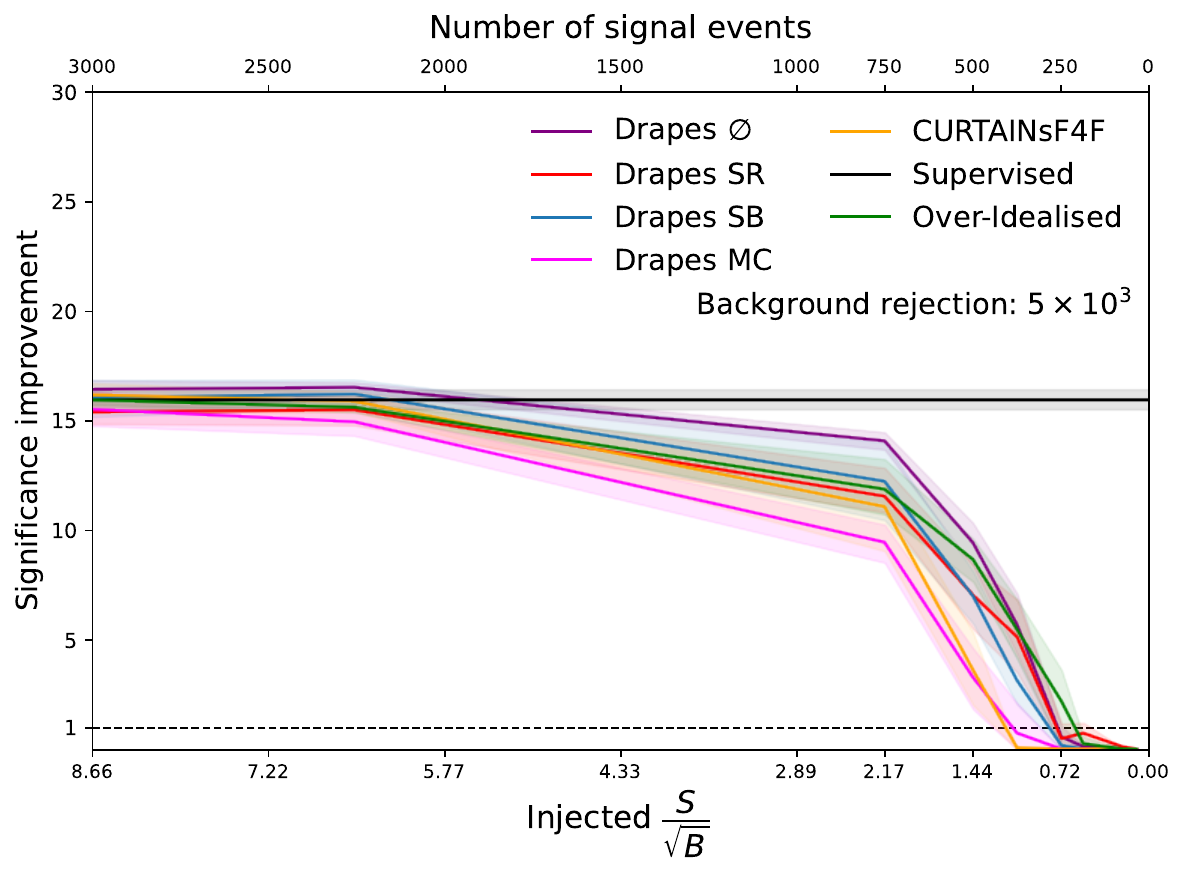}
    \end{subfigure}\\
    % \begin{subfigure}{0.48\textwidth}
    %     \centering
    %     \includegraphics[width=\textwidth]{figures/og_compare_rej_1000_sic_vs_nsig.pdf}
    % \end{subfigure}
    % \hfill
    % \begin{subfigure}{0.48\textwidth}
    %     \centering
    %     \includegraphics[width=\textwidth]{figures/og_compare_rej_5000_sic_vs_nsig.pdf}
    % \end{subfigure}
    \caption{Significance improvement at fixed background rejection rates of $10^3$ (left) and \mbox{$5 \times 10^3$} (right) as a function of the number of signal events in the signal region, \mbox{$3300\leq\mjj<3700$~GeV}, for \FfF~(orange), \drapesGen~(purple), \drapesSR~(red), \drapesSB~(blue), \drapesMC~(magenta), Supervised~(black), and Over-Idealised~(green).
    The injected significance measured from the number of signal and background events in the signal region before applying a cut $\frac{S}{\sqrt{B}}$ is also shown.
    The lines show the mean value of fifty independent classifiers, with the shaded band representing a 68\% uncertainty.
    A dashed line is shown at 1 to indicate where an improvement over the initial significance is achieved.}
    \label{fig:sic_vs_sig_all}
\end{figure}

In \cref{fig:sic_vs_sig_all} the significance performance at fixed background rejection rates is compared to \drapesGen for a range of signal values.
Here we see that \drapesSR comes closest to the overall performance of \drapesGen, with all methods becoming more similar as the injected $\rfrac{S}{\sqrt{B}}$ increases, though \drapesGen performs best for all values.
However, \drapesSR and \drapesSB remain competitive, with less than half the number of denoising steps as \drapesGen.

\subsection{\drapes with jet constituents}

When using low level inputs~(LLV) we focus on the two best drapes generation approaches, \drapesGen and \drapesSR.
We compare the performance to \drapes trained on high level variables~(HLV) and a supervised classifier trained on the same low level inputs.

\begin{figure}[htbp]
    \centering
    \begin{subfigure}{0.48\textwidth}
        \centering
        \includegraphics[width=\textwidth]{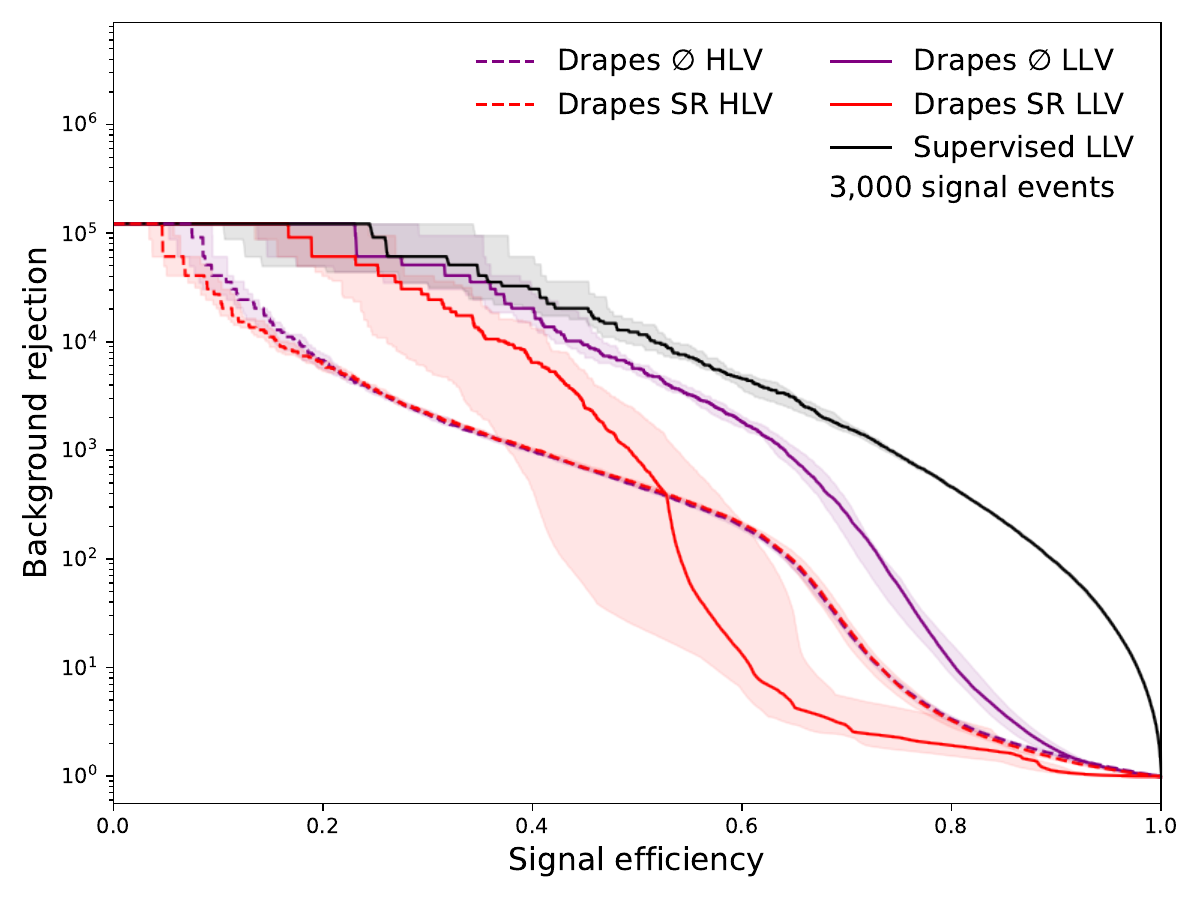}
    \end{subfigure}
    \hfill
    \begin{subfigure}{0.48\textwidth}
        \centering
        \includegraphics[width=\textwidth]{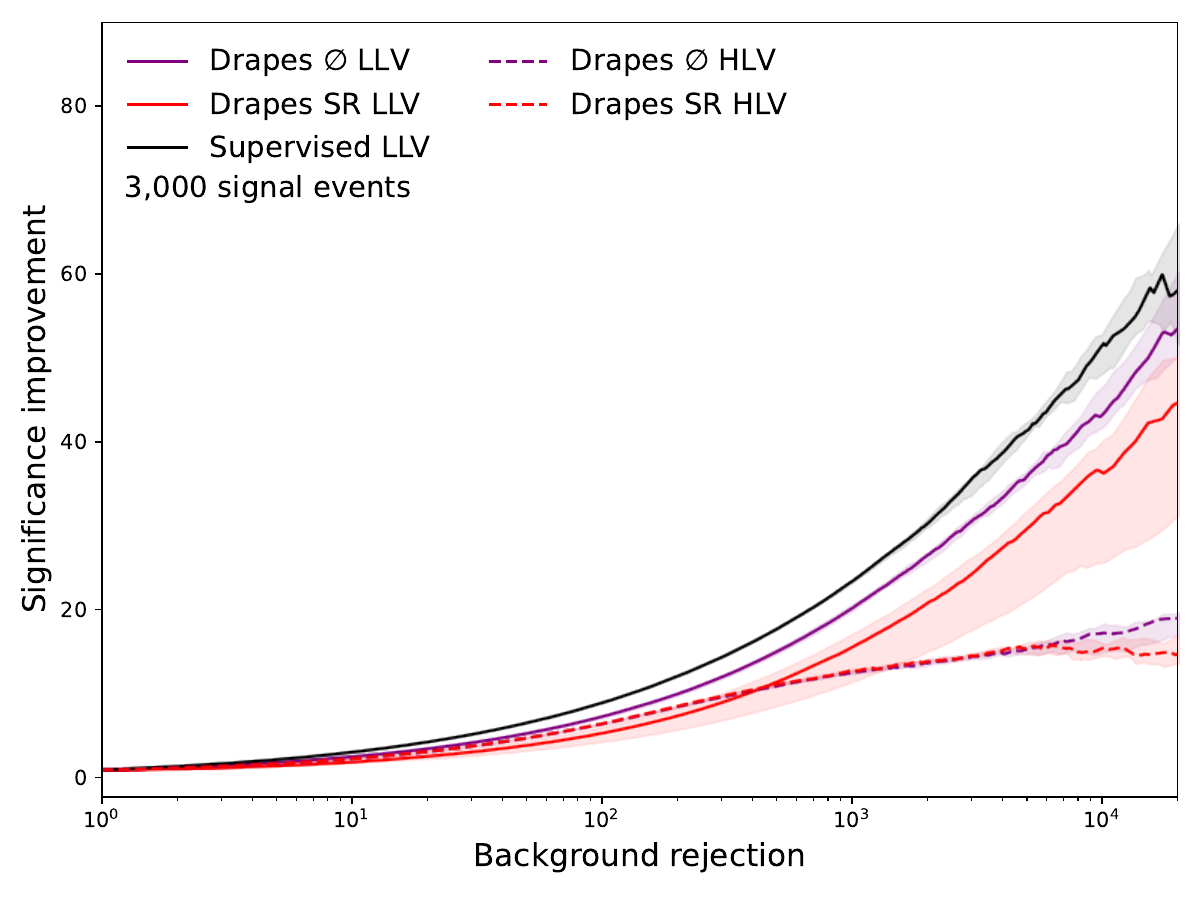}
    \end{subfigure}
    \caption{Background rejection as a function of signal efficiency (left) and significance improvement as a function of background rejection (right) for \drapesGen~(purple) and \drapesSR~(red).
    Models trained on the low level jet constituents are shown in a solid line (LLV), with models using the high level variables (HLV) are shown in a dashed line.
    All methods are trained on the sample with 3,000 injected signal events, and a signal region $3300\leq\mjj<3700$~GeV.
    The lines show the mean value of fifty independent classifiers, with the shaded band representing a 68\% uncertainty.
    A supervised classifier~(black) trained on low level jet constituents is shown for reference.}
    \label{fig:lowlevel_rocsic}
\end{figure}

\cref{fig:lowlevel_rocsic} shows the performance for the case of 3,000 injected signal events. 
We can see that \drapes trained on low level inputs substantially outperforms the high level features.
Here \drapesGen again results in overall better performance compared to partial diffusion, with \drapesSR suffering a non-negligible performance reduction not observed for HLV.
This is most notable in the ROC curve, where at high signal efficiencies \drapesSR~LLV has significantly lower background rejection.

\begin{figure}[htbp]
    \centering
    \begin{subfigure}{0.48\textwidth}
        \centering
        \includegraphics[width=\textwidth]{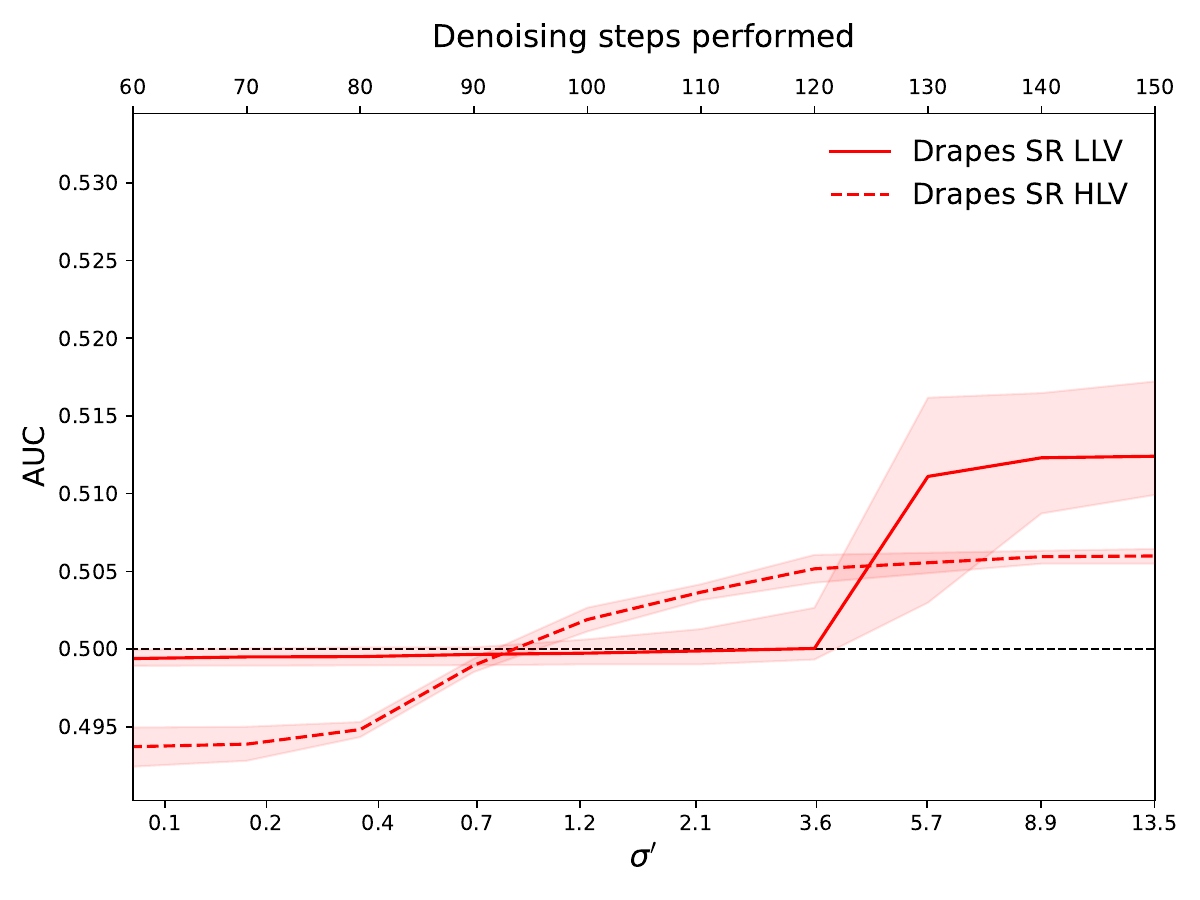}
    \end{subfigure}
    \hfill
    \begin{subfigure}{0.48\textwidth}
        \centering
        \includegraphics[width=\textwidth]{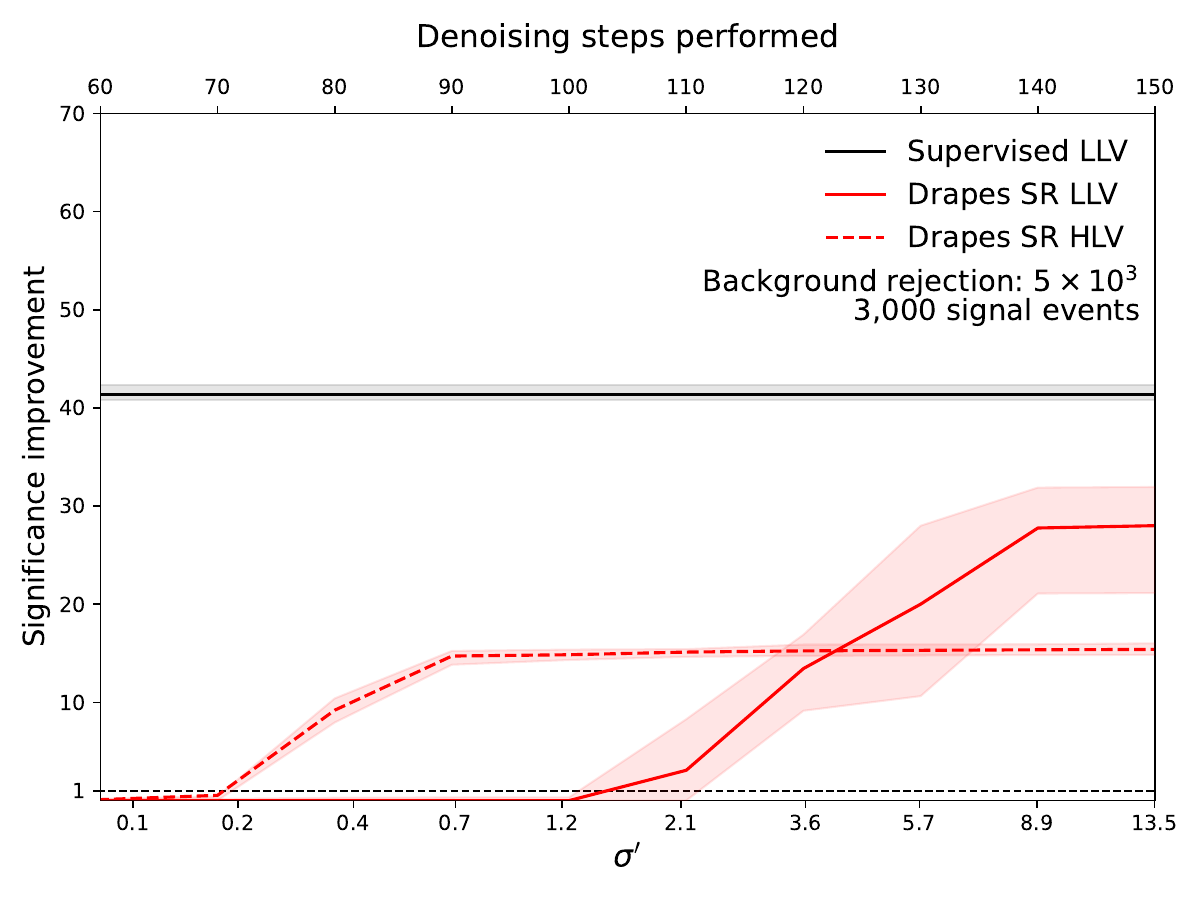}
    \end{subfigure}
    \caption{Impact on performance of varying the level of partial diffusion $\sigma^\prime$ in the case where there are no signal events (left) and 3,000 injected signal events (right).
    For the no signal case we show the AUC values of classifiers trained to separate the \drapesSR reference samples from true background data.
    For the case where signal events are present we show the significance improvement at a background rejection of 5,000.
    % Impact on the level of Background rejection as a function of signal efficiency (left) and significance improvement as a function of background rejection (right) for \drapesGen~(purple), \drapesSR~(red) and Supervised~(black).
    Models trained on the low level jet constituents are shown in a solid line (LLV), with models using the high level variables (HLV) are shown in a dashed line.
    All methods are trained on the sample with 3,000 injected signal events, and a signal region $3300\leq\mjj<3700$~GeV.
    The lines show the mean value of fifty independent classifiers, with the shaded band representing a 68\% uncertainty.
    For the significance improvement, a supervised classifier~(black) trained on low level jet constituents is shown for reference.}
    \label{fig:lowlevel_sigmap}
\end{figure}

For \drapesSR a value of $\sigma^\prime=8.9$ is used for LLV with $\sigma^\prime=0.62$ used for HLV as before.
The motivation for this can be seen in \cref{fig:lowlevel_sigmap}, where \drapesSR has little to no sensitivity below $\sigma^\prime=2$.
This suggests that the partial diffusion approaches have less direct benefit when applied to LLV in comparison to HLV.
However, this could be strongly dependent on how \drapesSR is applied when using the jet constituents.
In this work, the jet kinematics used to condition the \drapes model are unmodified from the signal region data, and partial diffusion is only applied to the jet constituents. 
Another option, which has not been studied in this work, would be to perform partial diffusion on both $p_4$ and the jet constituents.
This could improve the sensitivity of the partial diffusion method, and would be necessary for using partial diffusion with \drapesSB.
% However, for low values of $\sigma^\prime$ this could lead to discrepancies between $p_4$ and the initial jet constituents.
% It is not clear what impact this would have on the diffusion process.
% However, this could introduce discrepancies between the initial jet constituents and their associated jet four momenta.%model is used, with the jet kinematics taken directly from the signal region data.

\begin{figure}[htbp]
    \begin{subfigure}{0.48\textwidth}
        \centering
        \includegraphics[width=\textwidth]{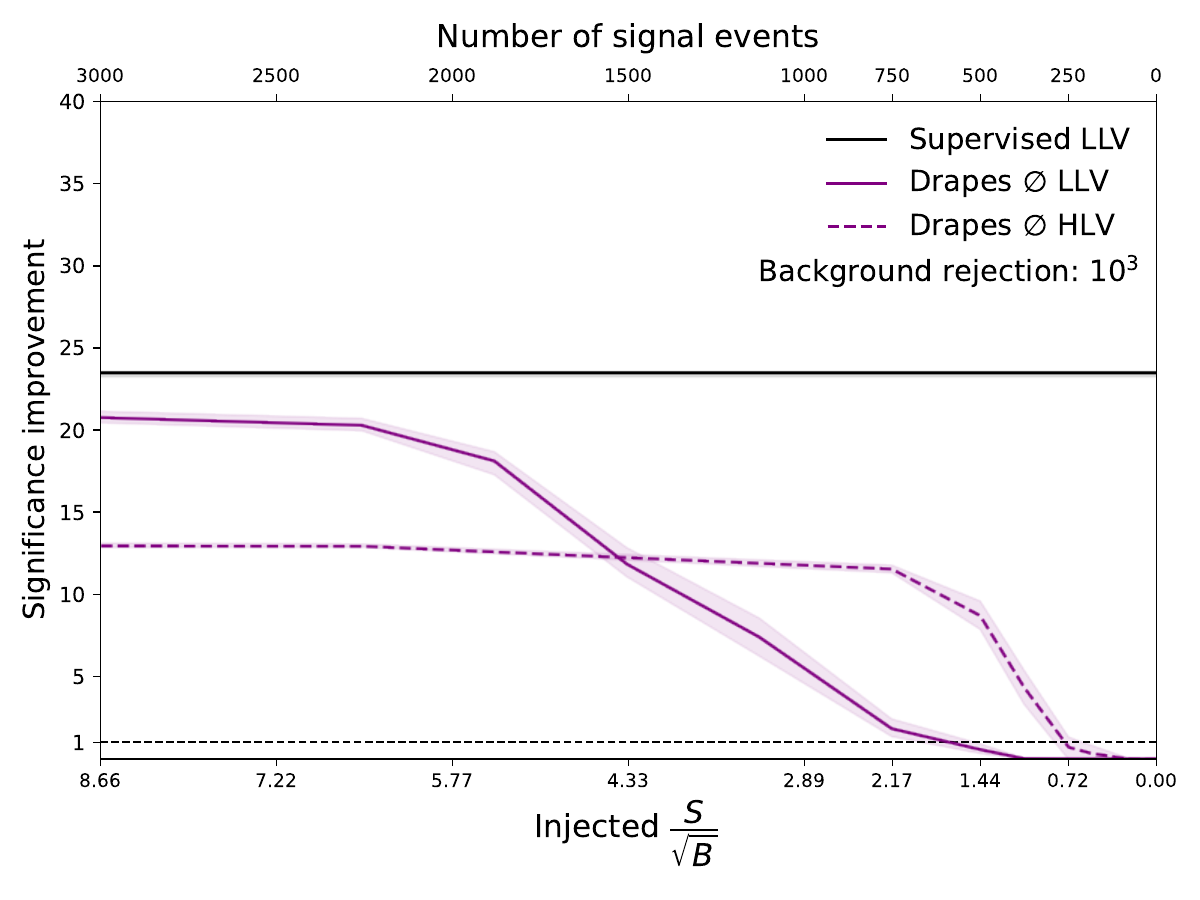}
    \end{subfigure}\hfill
    \begin{subfigure}{0.48\textwidth}
        \centering
        \includegraphics[width=\textwidth]{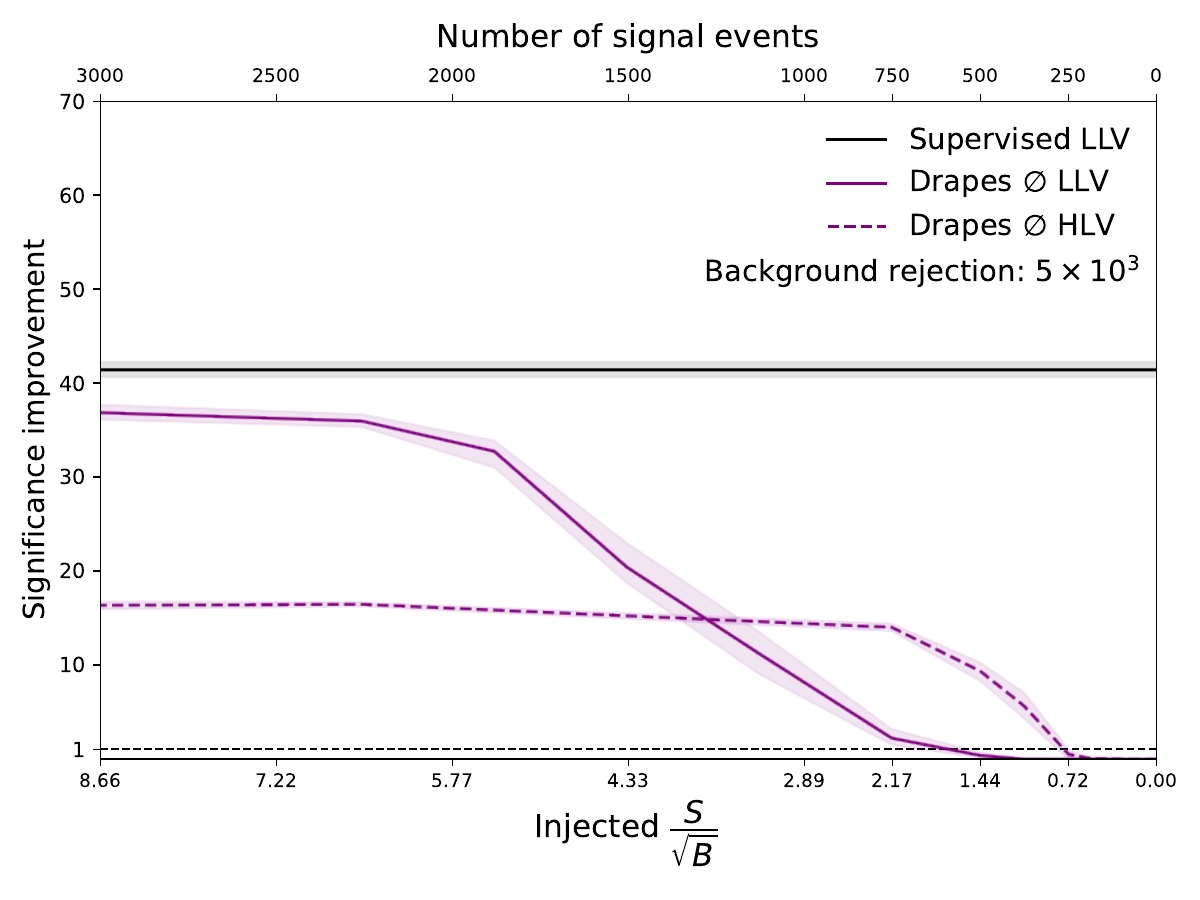}
    \end{subfigure}
    \caption{Significance improvement at fixed background rejection rates of $10^3$ (left) and \mbox{$5 \times 10^3$} (right) as a function of the number of signal events in the signal region, \mbox{$3300\leq\mjj<3700$~GeV}, for \drapesGen~(purple) using low level jet constituents (LLV, solid line) or high level variables (HLV, dashed line).
    The injected significance measured from the number of signal and background events in the signal region before applying a cut $(\frac{S}{\sqrt{B}})$ is also shown.
    The lines show the mean value of fifty independent classifiers, with the shaded band representing a 68\% uncertainty.
    A supervised classifier~(black) trained on low level jet constituents is shown for reference.}
    \label{fig:lowlevel_sic_vs_sig_all}
\end{figure}

The performance of \drapesGen~(LLV) as the amount of signal present changes is shown in \cref{fig:lowlevel_sic_vs_sig_all}.
Here we can see that although the performance achieved by \drapesGen~LLV substantially outperforms HLV where more signal is present, the performance drops off below 2,000 injected signal events.
\drapesGen trained on HLV outperforms the jet constituent approach in the region of interest where the injected significance is lower than 3.
This observation is consistent with what is observed in Ref.~\cite{Buhmann:2023acn}.
This drop in performance, however, can be expected and is reproducible in the idealised setting.
\CWoLa classifiers based on neural networks have been observed to lose sensitivity as the dimensionality of the input increases, especially with inputs which are not sensitive to signal processes~\cite{LaCathode,Finke:2023ltw,Freytsis:2023cjr}.
This effect is amplified as the number of signal events in the training data decreases.
In moving from five high level variables to jet constituents with \drapes, the dimensionality of the input to the weakly supervised neural networks has increased by more than a factor of 150.

% Part of this may lie in the observed sensitivity of \CWoLa with in

    \FloatBarrier
    \section{Conclusion}
    In this work we have introduced \drapes, a new approach for generating background reference samples for use in resonant anomaly searches, using diffusion models with deterministic samplers.
This approach can be used on both high level features as well as low level jet constituents.

For high level features, state-of-the-art performance is achieved on the LHCO R\&D dataset when using \drapes as a purely data-driven generative model.
Even better performance could be achieved by combining the generated background templates with boosted decision trees, rather than neural networks, for the \CWoLa classifier.

%%% Add in comment on low level
Furthermore, we demonstrate an advantage of the flexibility of diffusion models by applying \drapes to the generation of low level jet constituents.
This approach is more model agnostic than using high level variables, as it does not impose a choice of observables which may be sensitive to new physics processes.
Substantial improvements are observed for many levels of signal events.
However, for initial levels of signal significance below $4\sigma$ the performance degrades and is outperformed by the high level features.

In addition to pure template generation, we demonstrate the first application of partial diffusion in high energy physics.
Partial diffusion is a promising approach to generate new samples from an initial reference dataset, however on this dataset it does not reach the same level of performance as \drapesGen.
This is likely due to the nature of the application in high energy physics.
In other fields, anomalies can be characterised as single-sample out-of-distribution events.
In this case, distance based metrics between the original sample and the partially diffused sample can be used as an anomaly score.
However, in the case of resonant anomalies, we are searching for over-densities in feature space, localised in one observable (\mjj).
In this case, what is important is that we compare the differences in densities of our datasets.
With partial diffusion, if the initial samples are all in distribution, by performing the forward and reverse diffusion process, the initial density of the initial samples will be preserved.
This approach could still have merit when combined with another anomaly metric than the classifier based approaches studied here.
For example exploiting the displacement of samples in the signal region with \drapesSB, or assigning labels during training to side-band and signal region data which can be reversed at inference.

% In comparison to other methods \drapes also has additional advantages, for example the network architecture is far more flexible than what is allowed in normalizing flows, as applied in \CURTAINs and \CATHODE.
% This could open up additional possibilities for the feature space in resonant anomaly detection searches, or incorporating physics priors into the generative model.

    \section*{Acknowledgements}
    The authors would like to acknowledge funding through the SNSF Sinergia grant CRSII5\_193716 ``Robust Deep Density Models for High-Energy Particle Physics and Solar Flare Analysis (RODEM)''
and the SNSF project grant 200020\_212127 ``At the two upgrade frontiers: machine learning and the ITk Pixel detector''.
They would also like to acknowledge the funding from the Swiss Government Excellence Scholarships for Foreign Scholars.

    \bibliography{bib/rodem}
    \bibliographystyle{JHEP}
    
    \appendix
    % \section{Appendix}

    \section{Impact of side-band width}

As seen in Ref.~\cite{curtainsf4f}, the amount of data used to train the models producing reference background samples, has non-negligible impact on the performance of the \CWoLa classifiers.
This is strongly influenced by the widths of the side-bands around the signal region.
It can be beneficial to have a local training window around the signal region if it is expected that the background composition may not be homogenous across a wide range of \mjj.

Furthermore, by using narrower side-bands with the SR centred on the signal peak in \mjj, we can also study the impact of increasing the relative amount of signal data present in the training region without changing the data in the SR to train the \CWoLa classifier

\begin{figure}[hbpt]
    \centering
    \begin{subfigure}{0.48\textwidth}
        \centering
        \includegraphics[width=\textwidth]{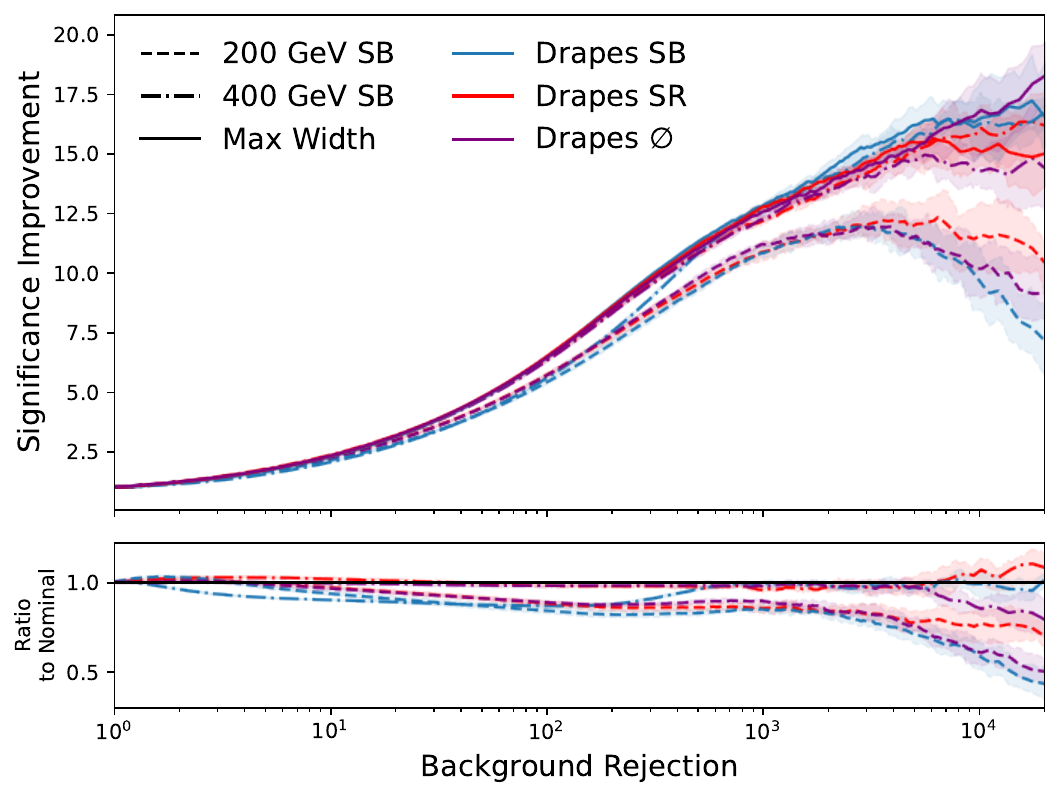}
    \end{subfigure}
    \hfill
    \begin{subfigure}{0.48\textwidth}
        \centering
        \includegraphics[width=\textwidth]{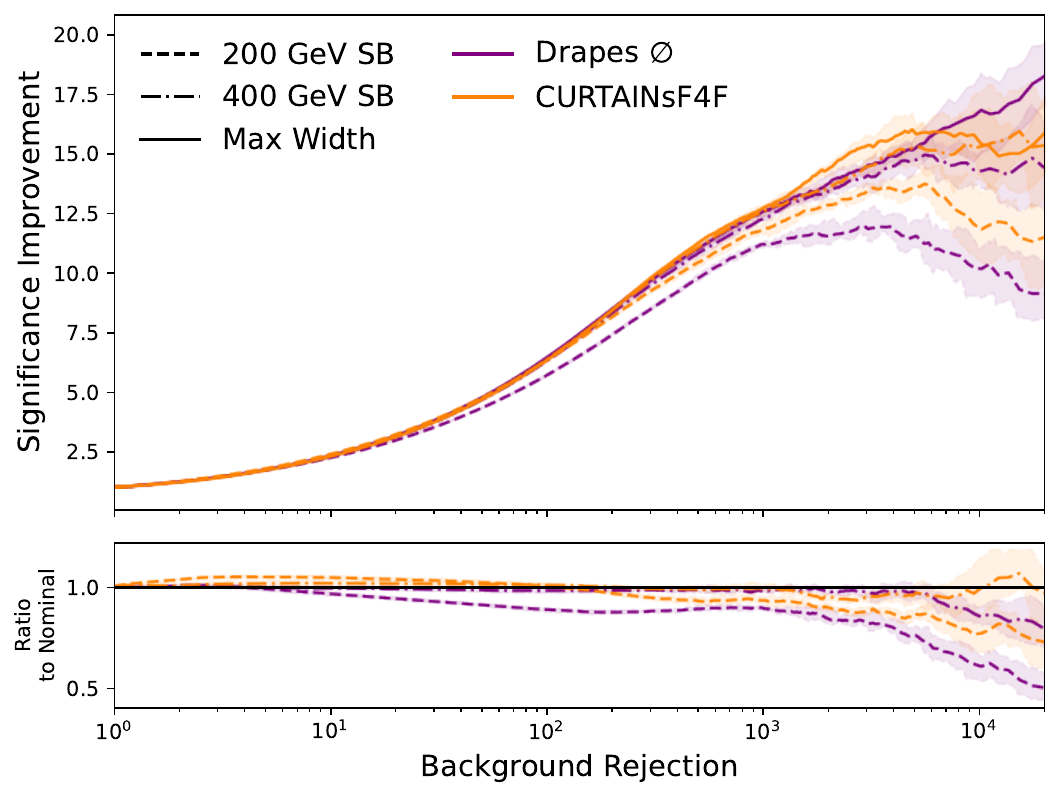}
    \end{subfigure}
    \caption{Significance improvement as a function of background rejection for \drapesGen~(purple), \drapesSR~(blue), \drapesSB~(red), and \FfF~(orange), trained on side-band data for three different side-band widths.
    % The diffusion model is trained using different side-band widths.
    % The `Max Width' case uses a side-band spanning 2.7 TeV  -- 6.0 TeV for \drapes, whereas for \FfF it is 2.8 TeV -- 4.5 TeV.
    % Both \drapesSR and \drapesSB are using initial samples with a noise rate of 0.6, whereas \drapesGen generates samples purely from noise.
    All methods are trained on the sample with 3,000 injected signal events, and a signal region $3300\leq\mjj<3700$~GeV.
    The lines show the mean value of fifty independent classifiers, with the shaded band representing a 68\% uncertainty.
    }
    \label{fig:sic_roc_width}
\end{figure}
% Therefore it is of interest to study how well the methods perform when restricted to a smaller subset of data by choosing narrower side-bands.
The impact of SB width on the performance of \drapes and \FfF are studied for two narrow side-bands in comparison to the nominal maximum width.
\cref{fig:sic_roc_width} show the performance for the case where there are 3,000 signal events in the data.
Here we see that all \drapes methods are more sensitive to the side-band width than \FfF, demonstrating similar behaviour to \CATHODE observed in Ref.~\cite{curtainsf4f}.
However, \drapesSR retains slightly more of its performance at high background rejection values.
% Here we see for \drapes that all generation methods are equally sensitive to the side-band width, though \drapesSB retains slightly more .
% Furthermore, we see that these methods are more sensitive to the side-band width than \FfF, demonstrating similar behaviour to \CATHODE observed in Ref.~\cite{curtainsf4f}.

\section{Studying oversampling}
\subsection{Impact of oversampling background template}

\begin{figure}[hbpt]
    \centering
    \begin{subfigure}{0.48\textwidth}
        \centering
        \includegraphics[width=\textwidth]{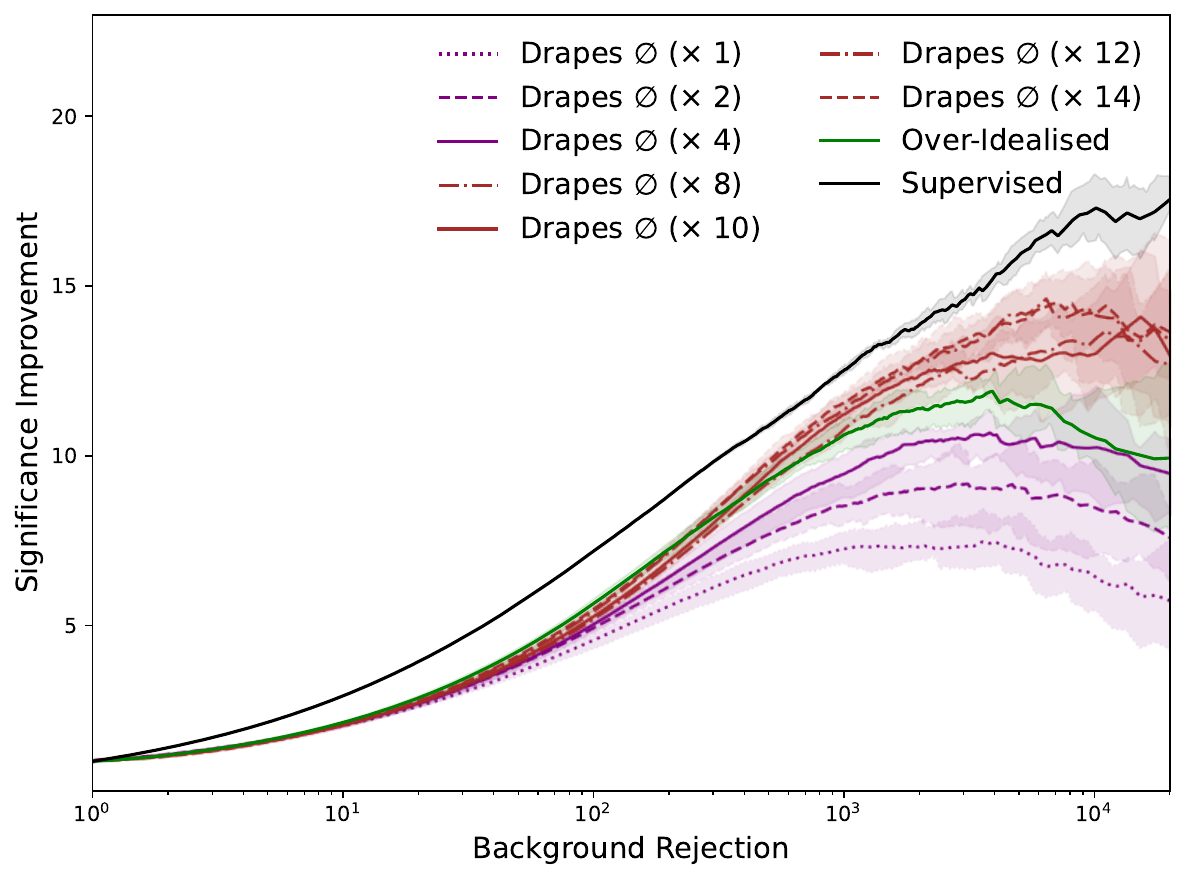}
    \end{subfigure}
    \hfill
    \begin{subfigure}{0.48\textwidth}
        \centering
        \includegraphics[width=\textwidth]{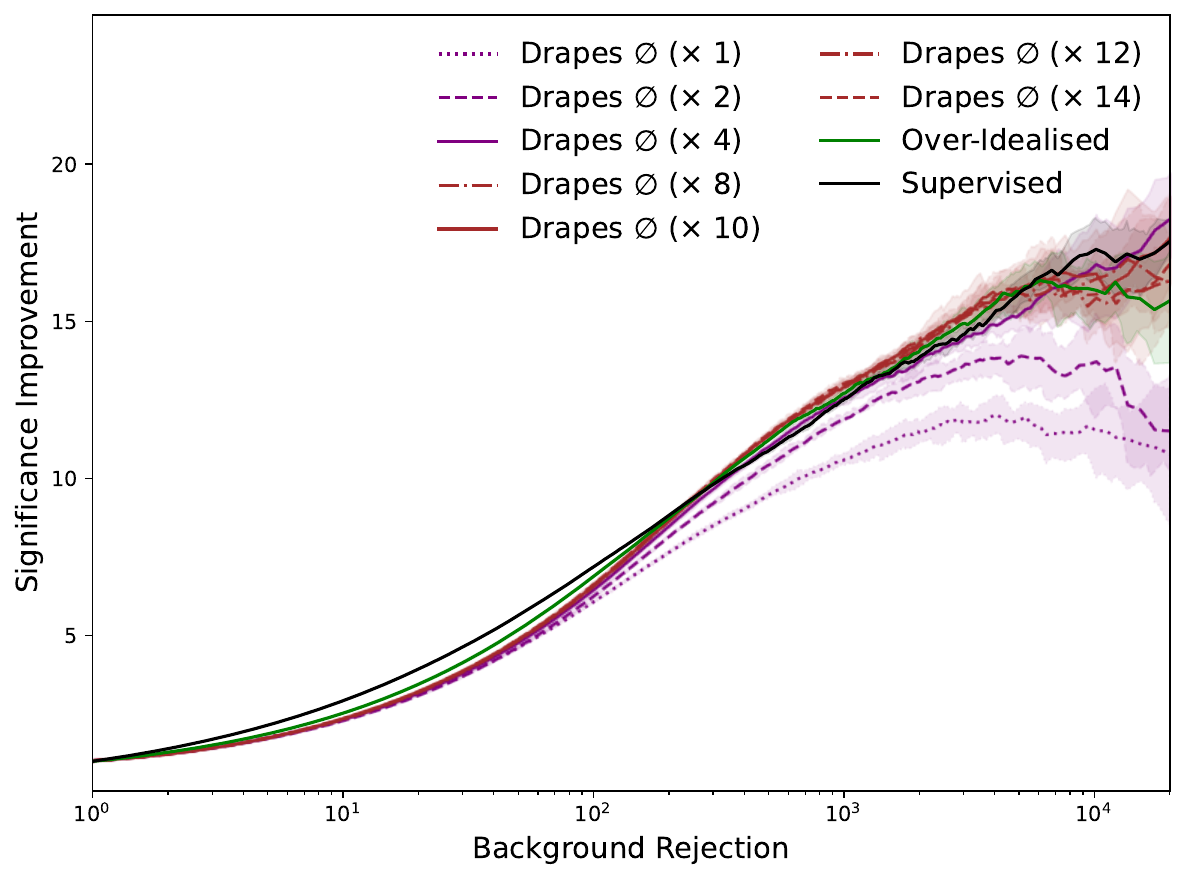}
    \end{subfigure}
    \caption{Significance improvement as a function of background rejection for \drapesGen for different levels of background oversampling, compared to Over-Idealised~(green) and Supervised~(black).
    All methods are trained on the sample with 1,000 injected signal events (left) and 3,000 injected signal events (right), and a signal region $3300\leq\mjj<3700$~GeV.
    The lines show the mean value of fifty independent classifiers, with the shaded band representing a 68\% uncertainty.
    }
    \label{fig:sic_vs_sig_oversample}
\end{figure}

In \cref{fig:sic_vs_sig_oversample} the impact of oversampling the background reference samples with \drapesGen i studied for 1,000 and 3,000 injected signal events.
We observe that a saturation of performance is achieved at a factor of four for 3,000 signal events, however in the case of 1,000 signal events more training data are required to saturate the performance.
Here saturation is achieved at a factor of 12.
For reference, the over-idealised curve has an oversampling factor of five.

\subsection{Oversampling signal region data}
With \drapes we observe that oversampling the background template is able to bring substantial improvement to the overall performance of the \CWoLa classifier.
This is something which has also been observed and exploited in many other approaches~\cite{cathode,curtains,LaCathode,feta,curtainsf4f,interplay}, and also brings performance gains in the idealised case.
However, this oversampling is only performed on one class in the classifier, with the real signal region data untouched.

Another observation is that even in the idealised case that the performance degrades as the number of signal events present in the signal region decreases.
By training a \drapes model on the data in the signal region, we can learn the conditional density for the combination of both signal and background events as a function of \mjj.
Using this model, \genSR, we can sampling multiple events for each value of \mjj in the signal region and see is able to improve the sensitivity of the \CWoLa classifiers in the case of few signal events.
The \CWoLa classifiers are subsequently trained to separate this oversampled signal region data from the oversampled \drapes background reference samples. 
Using generative models to inflate training statistics has been observed to improve the performance relative to limited available statistics~\cite{Butter:2020qhk,Bieringer:2022cbs}.
However, to our knowledge this is the first test of this observation in the context of weakly supervised learning.

\begin{figure}[hbpt]
    \centering
    \begin{subfigure}{0.48\textwidth}
        \centering
        \includegraphics[width=\textwidth]{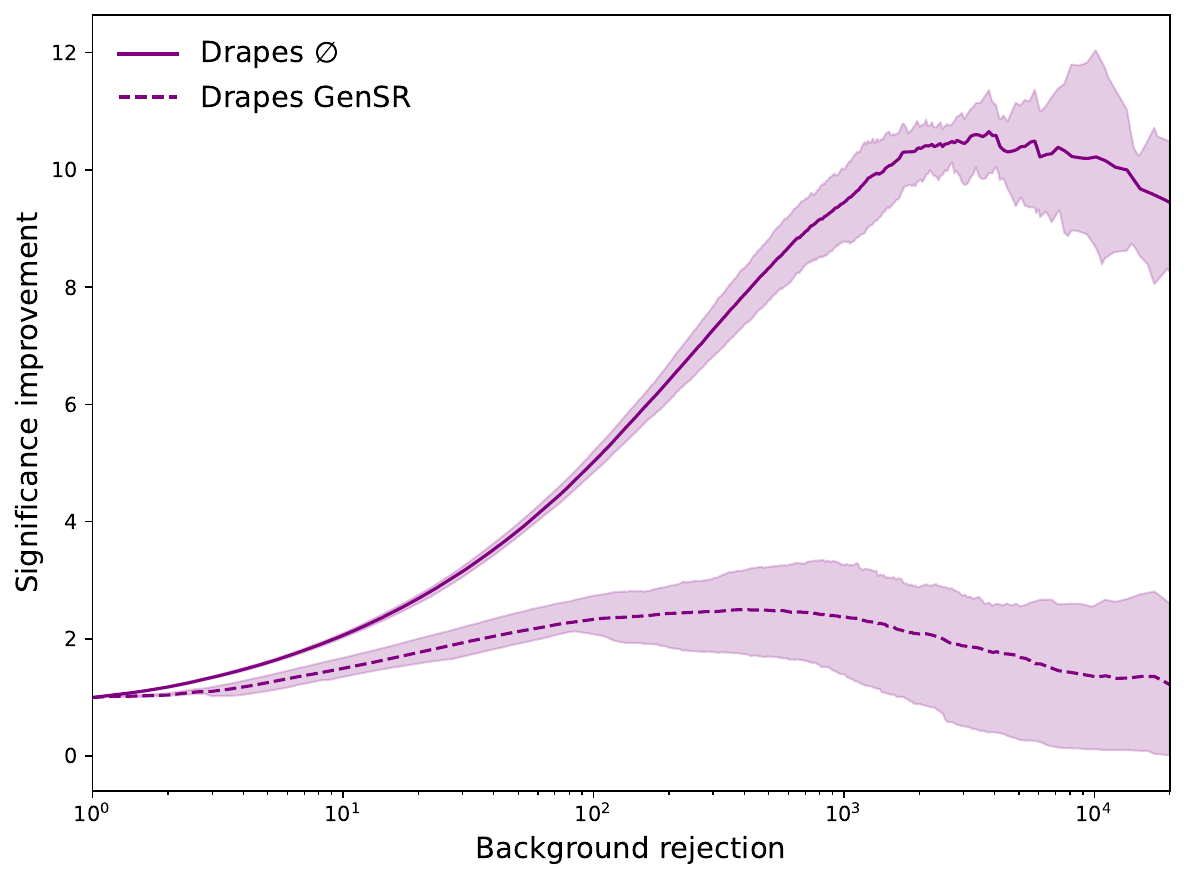}
    \end{subfigure}
    \hfill
    \begin{subfigure}{0.48\textwidth}
        \centering
        \includegraphics[width=\textwidth]{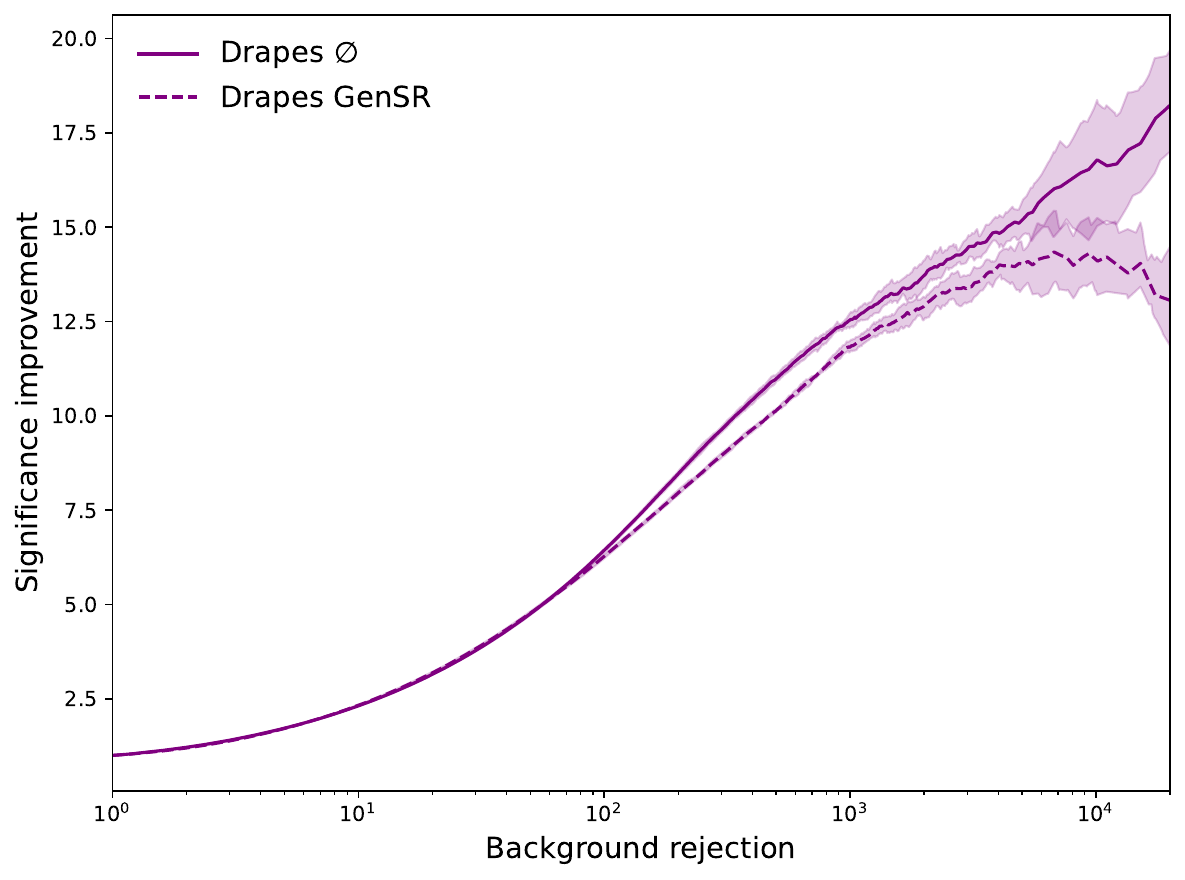}
    \end{subfigure}

    \caption{Significance improvement as a function of background rejection for 1,000 (left) and 3,000 (right) injected signal events for the nominal \drapesGen approach~(purple, solid) and when \drapes~GenSR is used to oversample the signal region data~(purple, dashed).
    The lines show the mean value of fifty independent classifiers, with the shaded band representing a 68\% uncertainty.
    }
    \label{fig:genSR}
\end{figure}

To study this we compare the nominal \drapesGen against a setup in which a second \drapesGen model has been trained inclusive of the signal region, and is used to oversample the signal region data.
The SB data is also included when training the new model, labelled \drapes~GenSR, to minimise the chance of the \CWoLa classifier only identifying differences in the background modelling arising from different training statistics.
Unfortunately, in \cref{fig:genSR} we find that the classifiers trained using \drapes~GenSR perform worse than when using the real signal region data.
This is especially the case for low amounts of injected signal, and suggests the \drapes model is not able to sufficiently capture the distribution over signal events when the training dataset comprises mostly background data, and at higher amounts of signal only approaches the upper limit of performance already reached with \drapesGen.

% By training on the With a diffusion model, or equally a normalizing flow, we can hope to learn the conditional density for both signal and background events with the correct rate as a function of \mjj.
% Using the same procedure as \drapesGen we train a new diffusion model, dubbed \genSR, with which for each value of \mjj we can now generate multiple events.
% \clearpage

\section{Additional figures}

\subsection{Partial diffusion}

\begin{figure}[hbpt]
    \centering
    \includegraphics[width=0.4\textwidth]{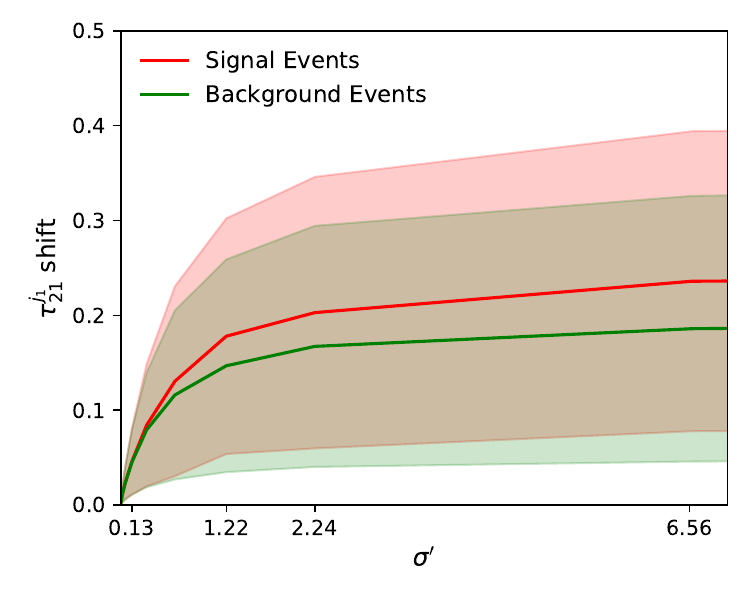}
    \includegraphics[width=0.4\textwidth]{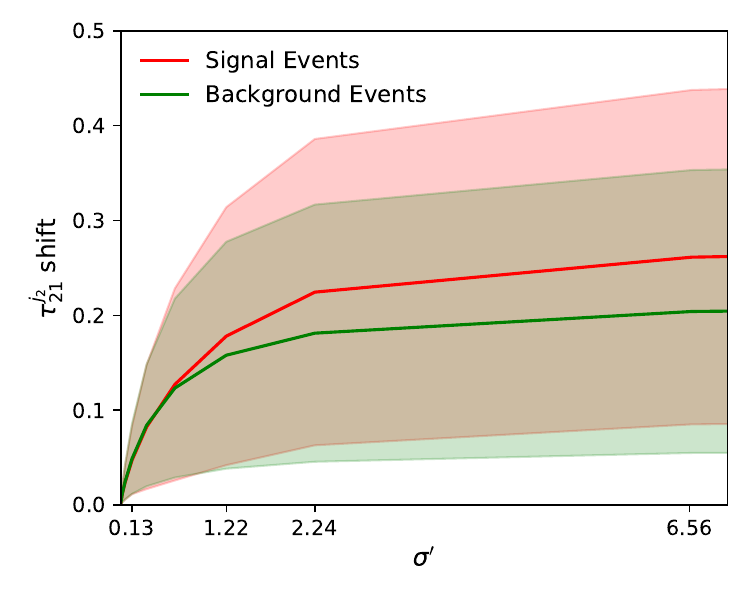}
    \includegraphics[width=0.4\textwidth]{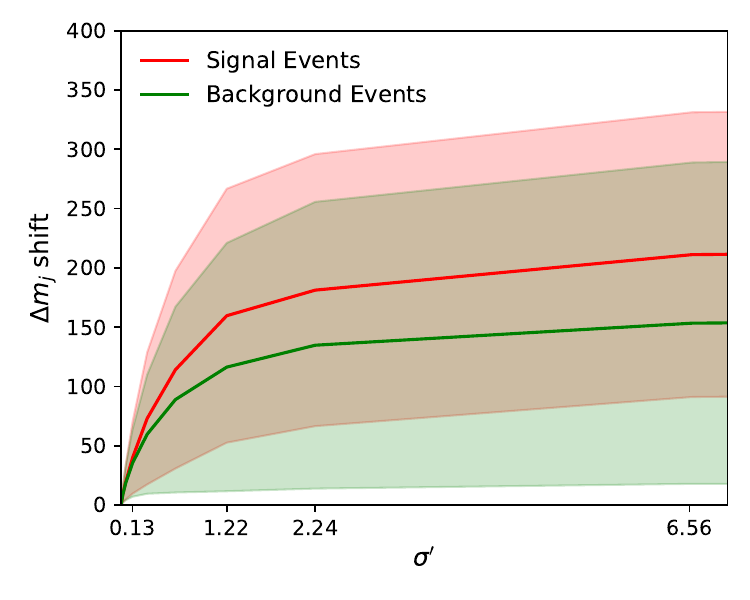}
    \includegraphics[width=0.4\textwidth]{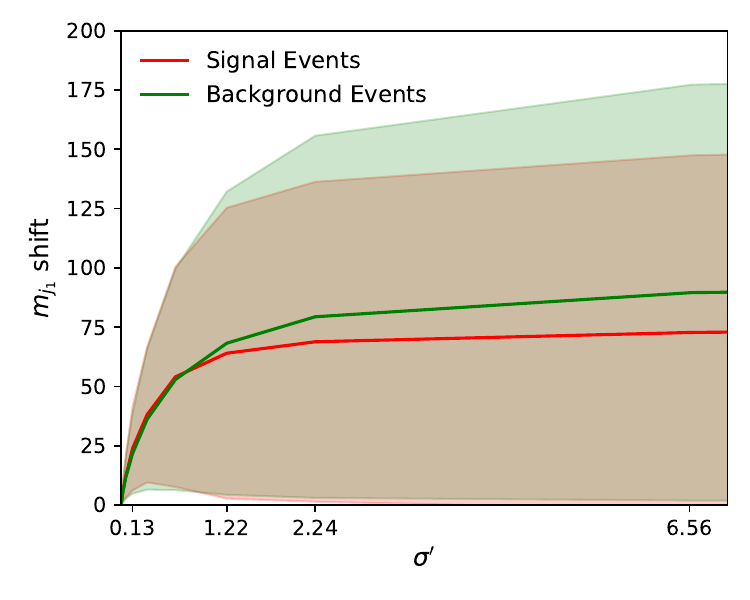}
    \includegraphics[width=0.4\textwidth]{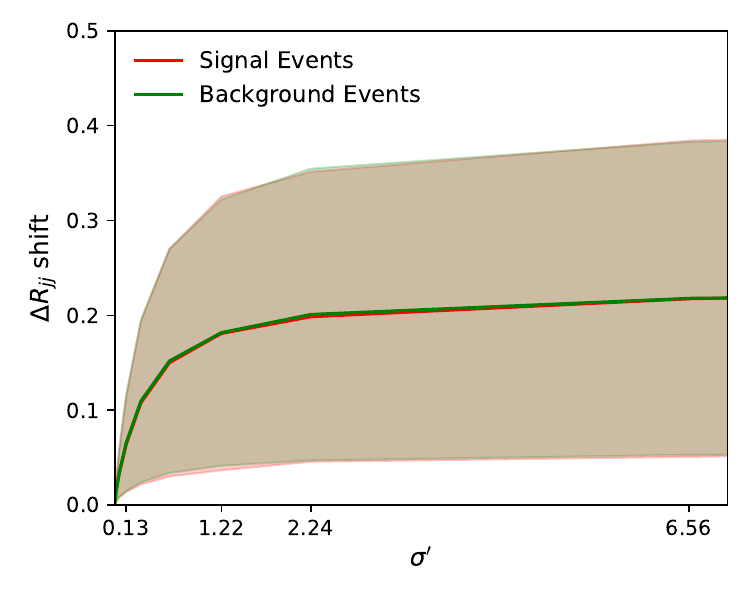}
    \caption{Displacement of input jets in the signal region after adding then removing noise with various noise rates using \drapesSR.
    A noise rate of 0 represents the input data without any noise perturbation or diffusion steps, and a rate of 80 corresponds to the case where the inputs are completely noise.
    The same diffusion model is used which was trained on the sample with 3,000 injected signal events.
    }
    \label{fig:shift_noise}
\end{figure}

When using partial diffusion noise is added to initial events which are subseqeuntly denoised.
The mean absolute shift between the input data and denoised output data is shown as a function of $\sigma^\prime$ in \cref{fig:shift_noise}.
In other domains, this displacement can be used as a single event classification score, with out of distribution often resulting in larger displacements.
However, as can be seen for $m_{j_1}$, the signal events have a smaller displacement for all values of $\sigma^\prime$.
% This is due to the more localised distribution for signal events, leading to a 

\begin{figure}[hbpt]
    \centering
    \includegraphics[width=0.32\textwidth]{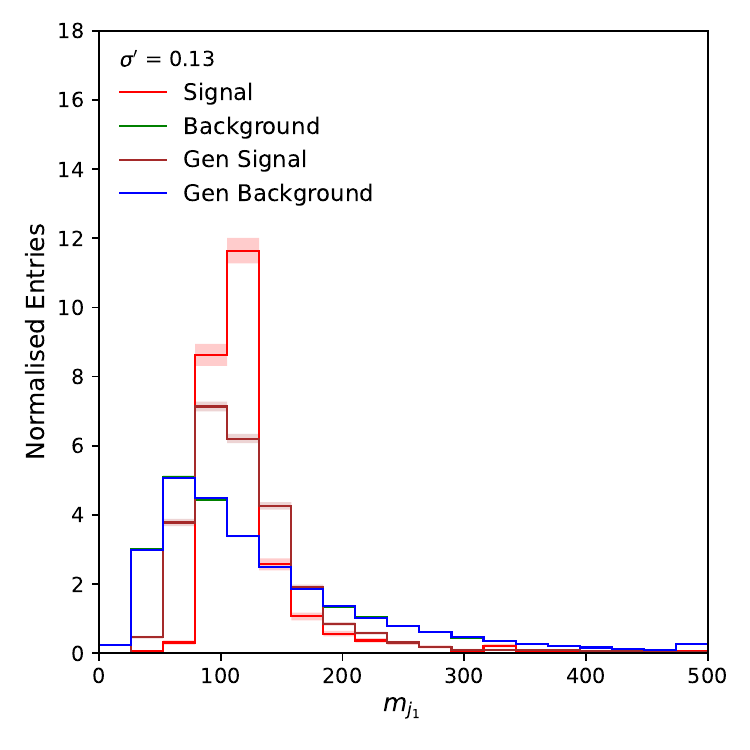}
    \includegraphics[width=0.32\textwidth]{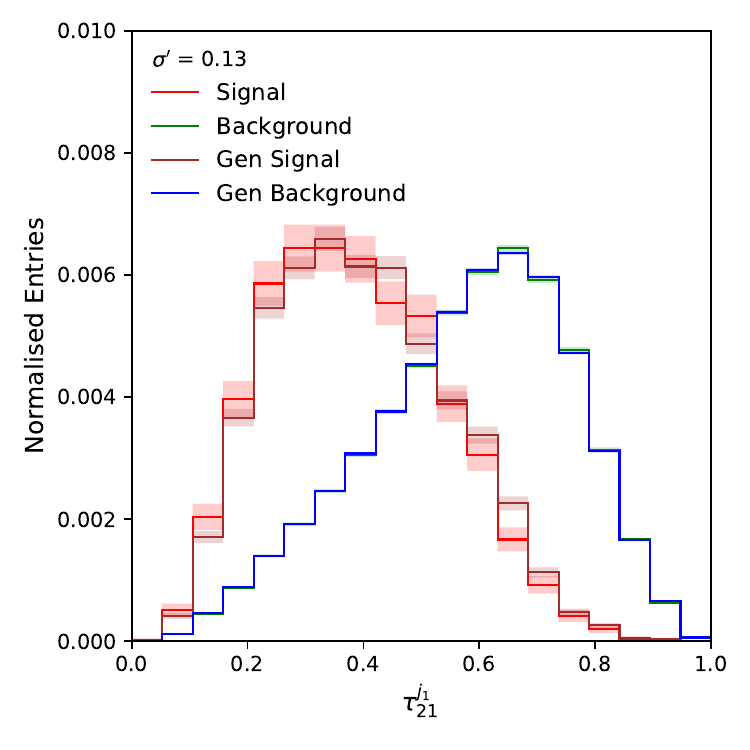}
    \includegraphics[width=0.32\textwidth]{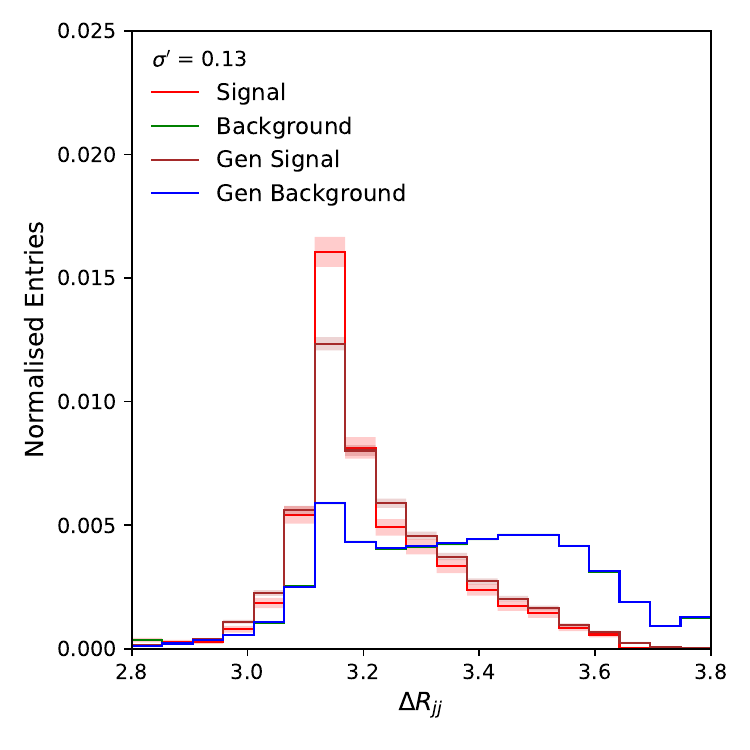}\\
    \includegraphics[width=0.32\textwidth]{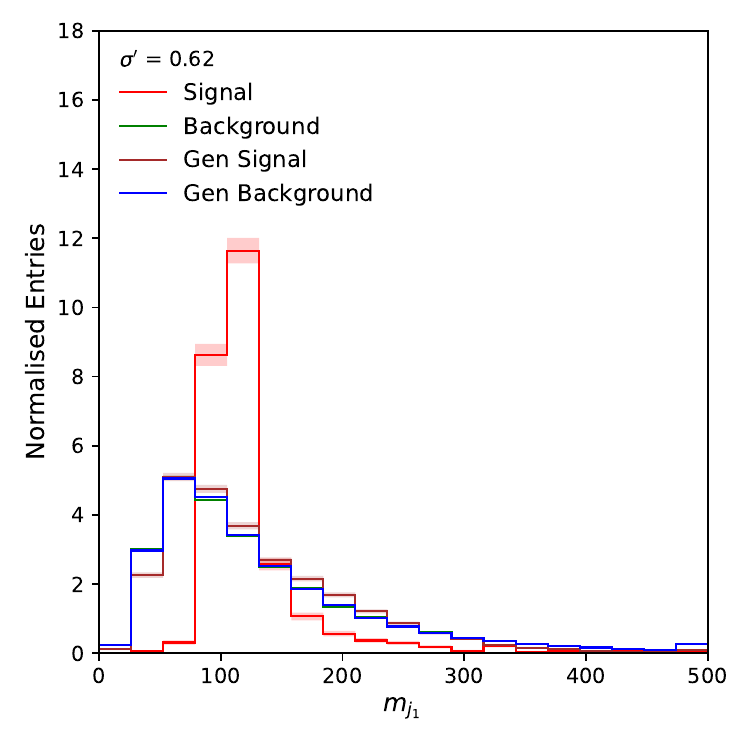}
    \includegraphics[width=0.32\textwidth]{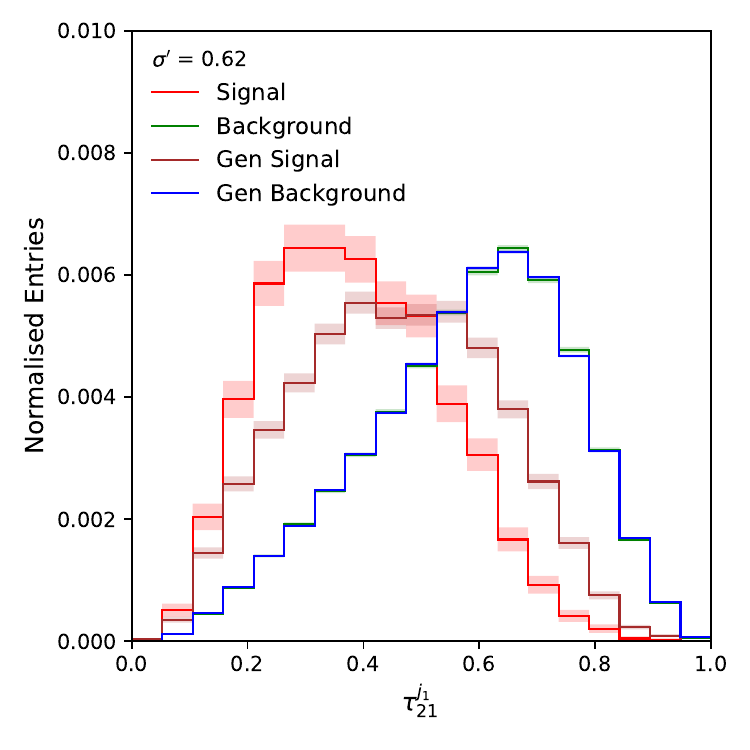}
    \includegraphics[width=0.32\textwidth]{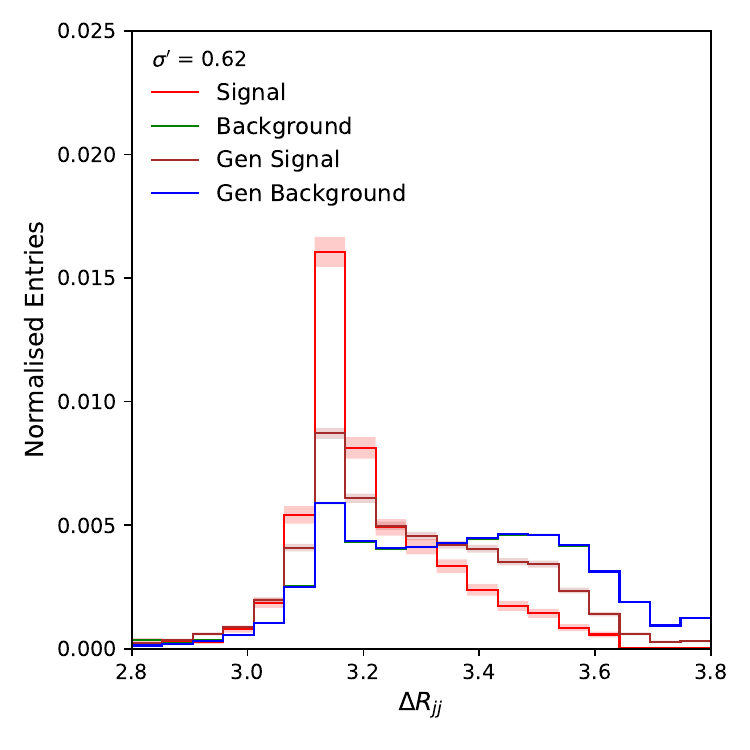}\\
    \includegraphics[width=0.32\textwidth]{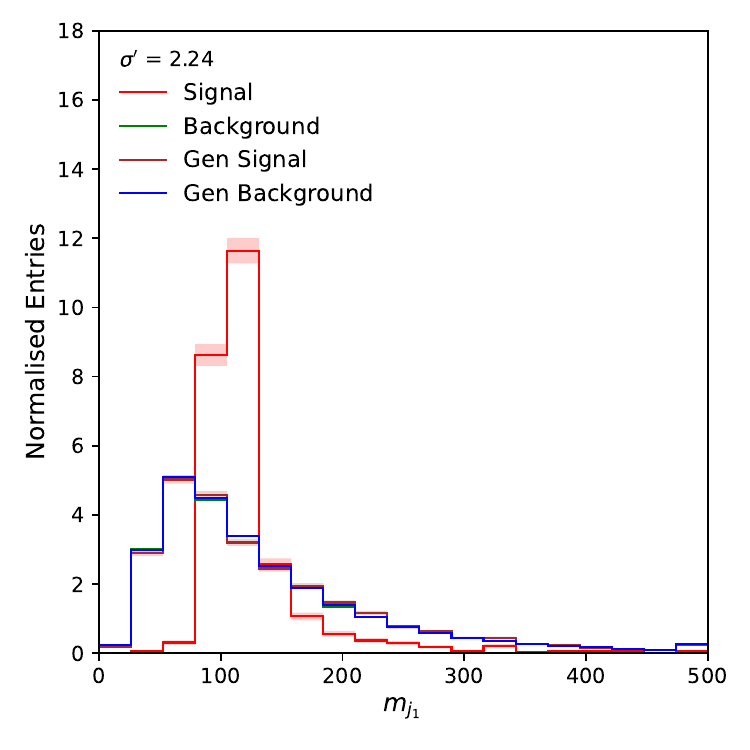}
    \includegraphics[width=0.32\textwidth]{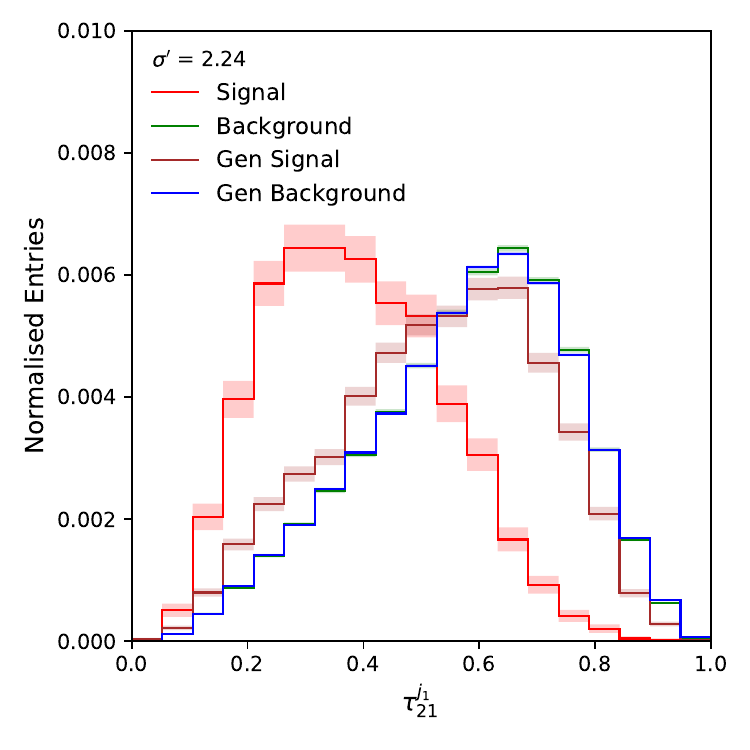}
    \includegraphics[width=0.32\textwidth]{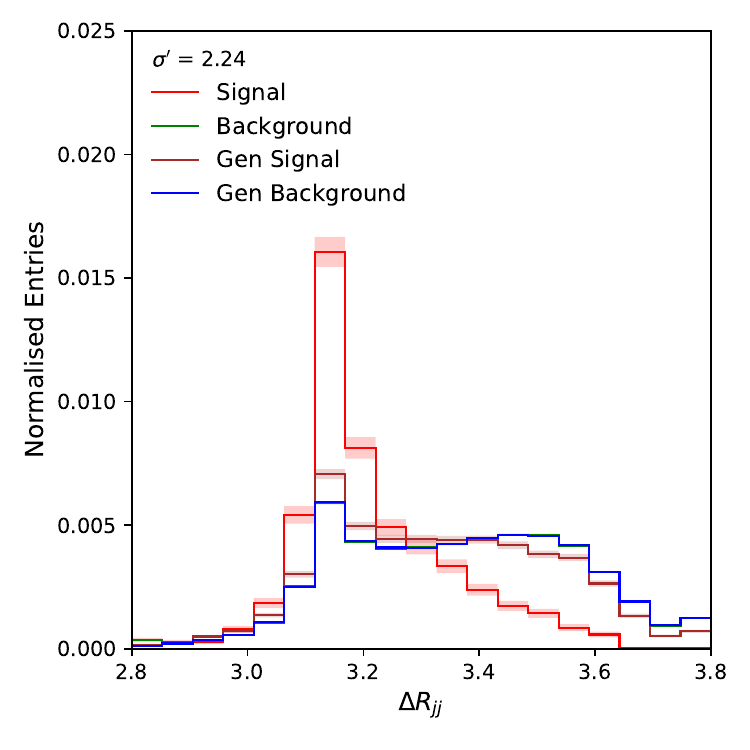}\\
    \includegraphics[width=0.32\textwidth]{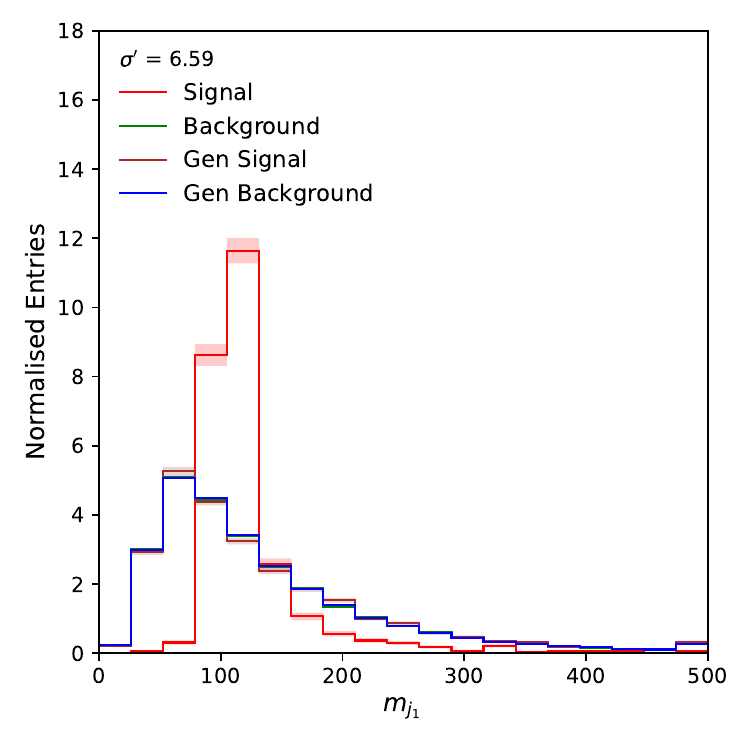}
    \includegraphics[width=0.32\textwidth]{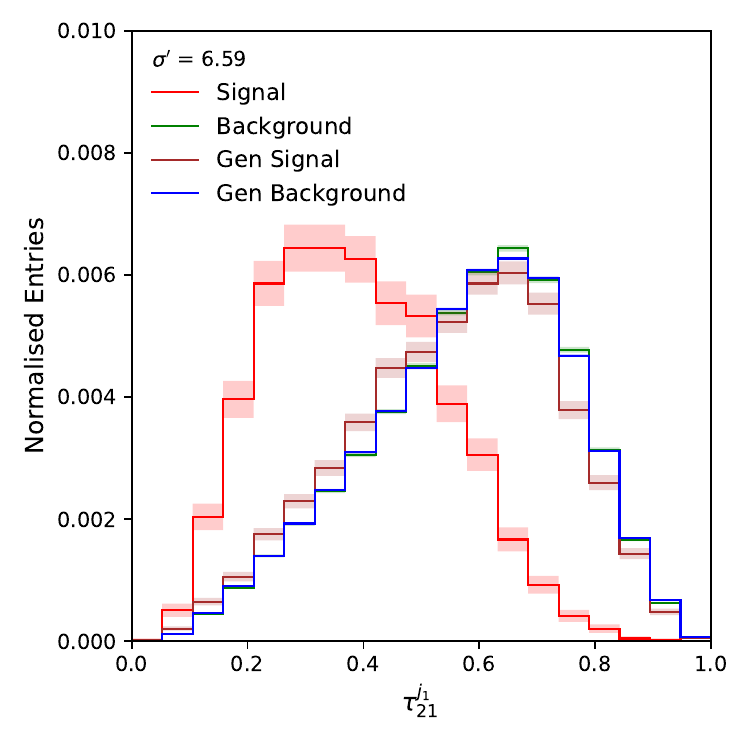}
    \includegraphics[width=0.32\textwidth]{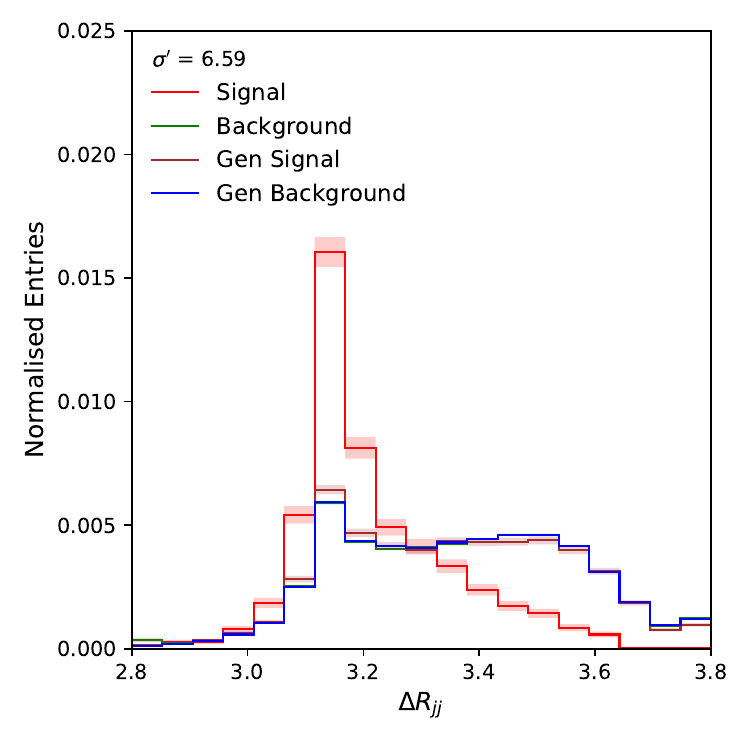}\\
    \caption{Events generated with \drapesSR for increasing values of $\sigma^\prime$ for $m_J^1$, $\tau_{21}^{j_1}$ and $\Delta R_{jj}$, shown separately for the case where initial events are from the signal (Gen Signal, brown) and background (Gen Background, blue) processes.
    The true signal (red) and background (green) distributions are shown for reference.}
    \label{fig:dist_noise}
\end{figure}

In \cref{fig:dist_noise} distributions for signal and background data after applying the forward and inverse diffusion processes in partial diffusion are compared to the true signal and background distributions.
In all cases, background events result in generated data following the same ground truth distribution.
However, in the case of signal events, as $\sigma^\prime$ increases, the resulting distributions become more background like, however even at $\sigma^\prime=2.24$ there remain differences between the generated data using signal as a starting point, and the background true distribution.

% \begin{figure}[hbpt]
%     \centering
%     \includegraphics[width=0.32\textwidth]{figures/dist/del_R_0.1.pdf}
%     \includegraphics[width=0.32\textwidth]{figures/dist/del_R_0.3.pdf}
%     \includegraphics[width=0.32\textwidth]{figures/dist/del_R_0.4.pdf}
%     \includegraphics[width=0.32\textwidth]{figures/dist/del_R_0.5.pdf}
%     \includegraphics[width=0.32\textwidth]{figures/dist/del_R_0.6.pdf}
%     \includegraphics[width=0.32\textwidth]{figures/dist/del_R_1.0.pdf}
%     \caption{Change in the distribution of the $\Delta R_{jj}$ variable as a function of the noise rate when using \drapesSR.
%     }
%     \label{fig:dist_noise}
% \end{figure}

\begin{figure}[hbpt]
    \centering
    \begin{subfigure}{0.48\textwidth}
        \centering
        \includegraphics[width=\textwidth]{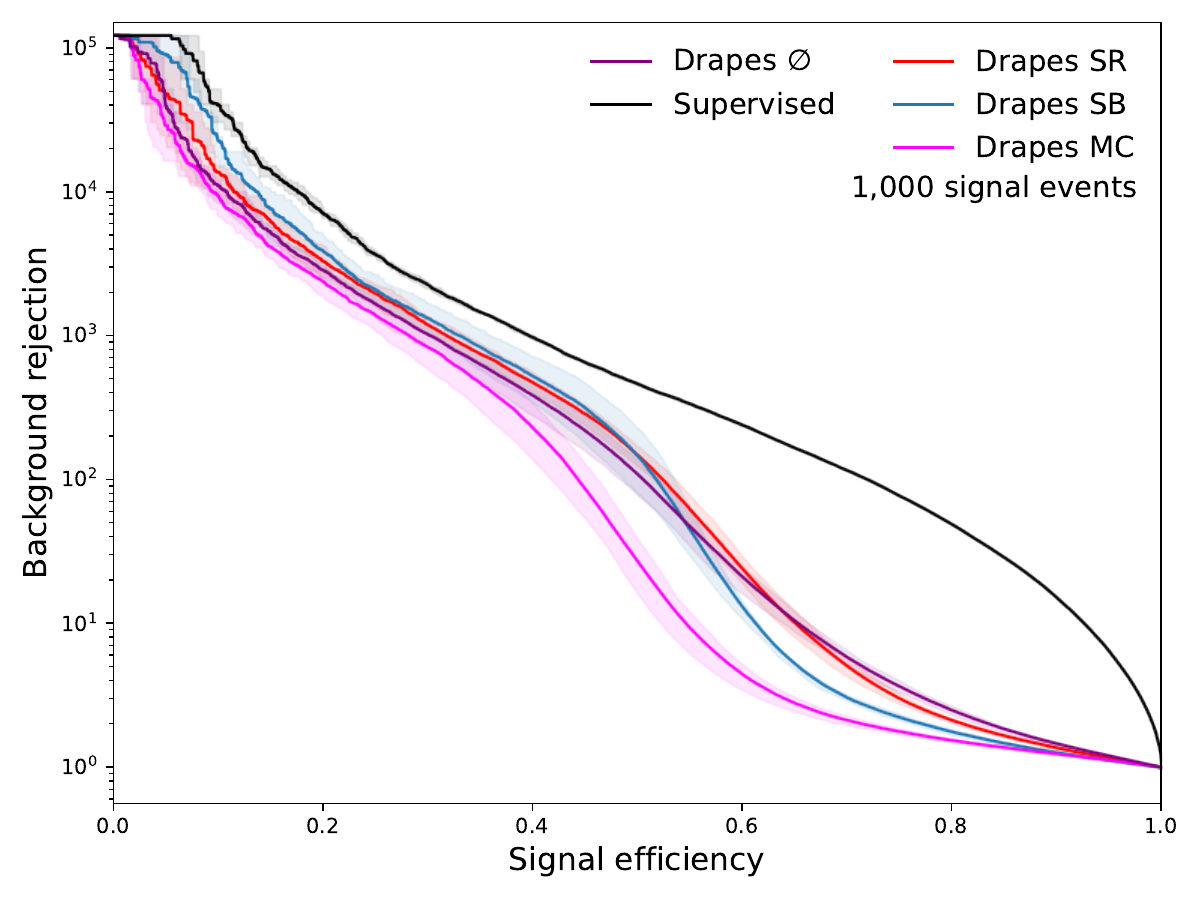}
    \end{subfigure}
    \hfill
    \begin{subfigure}{0.48\textwidth}
        \centering
        \includegraphics[width=\textwidth]{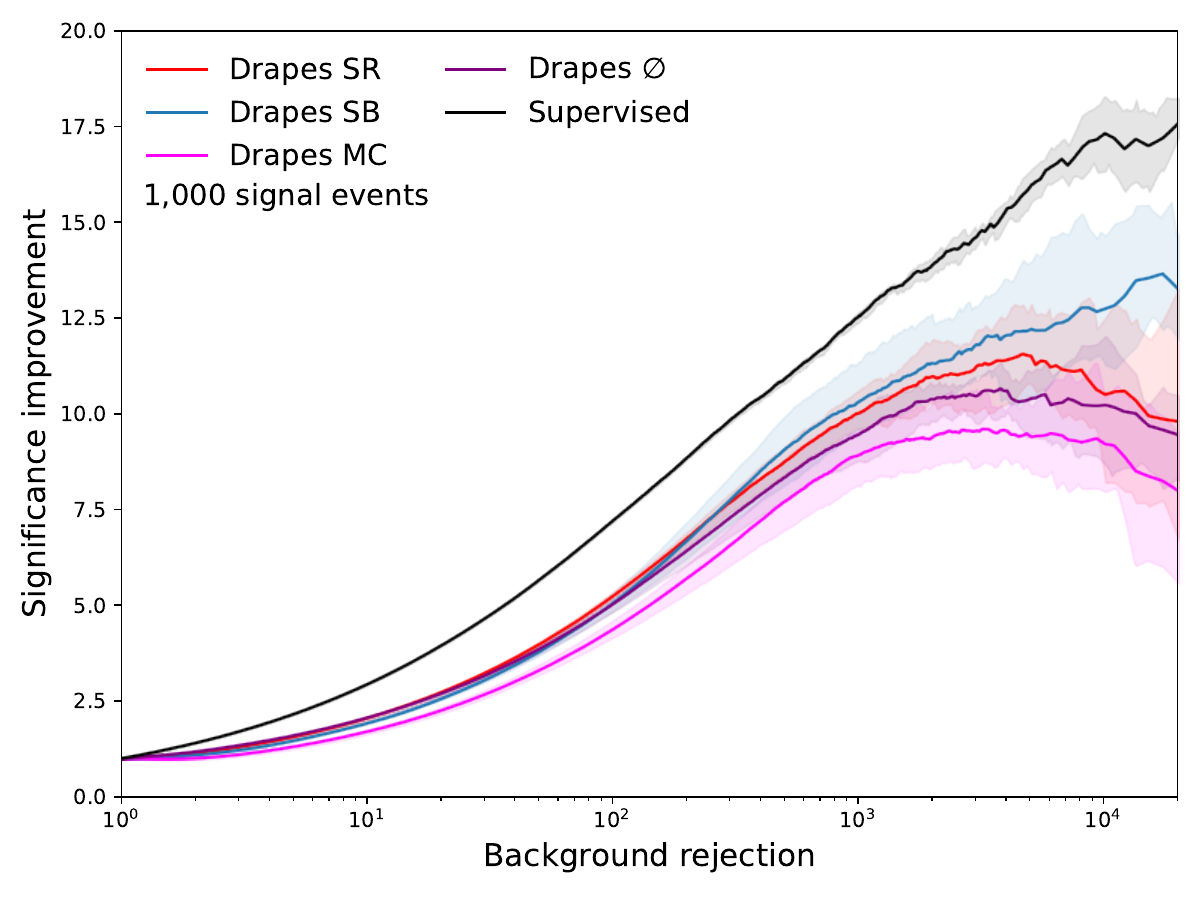}
    \end{subfigure}
    \begin{subfigure}{0.48\textwidth}
        \centering
        \includegraphics[width=\textwidth]{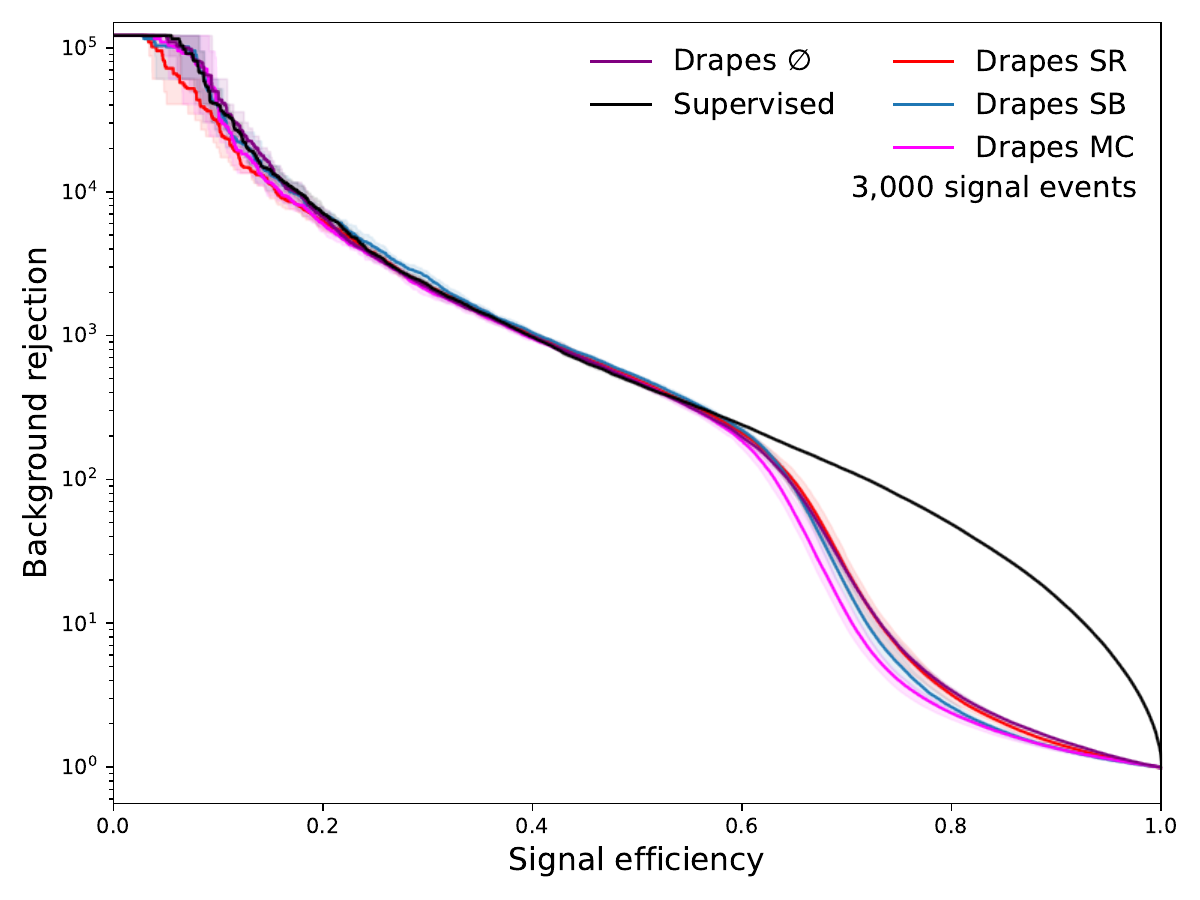}
    \end{subfigure}
    \hfill
    \begin{subfigure}{0.48\textwidth}
        \centering
        \includegraphics[width=\textwidth]{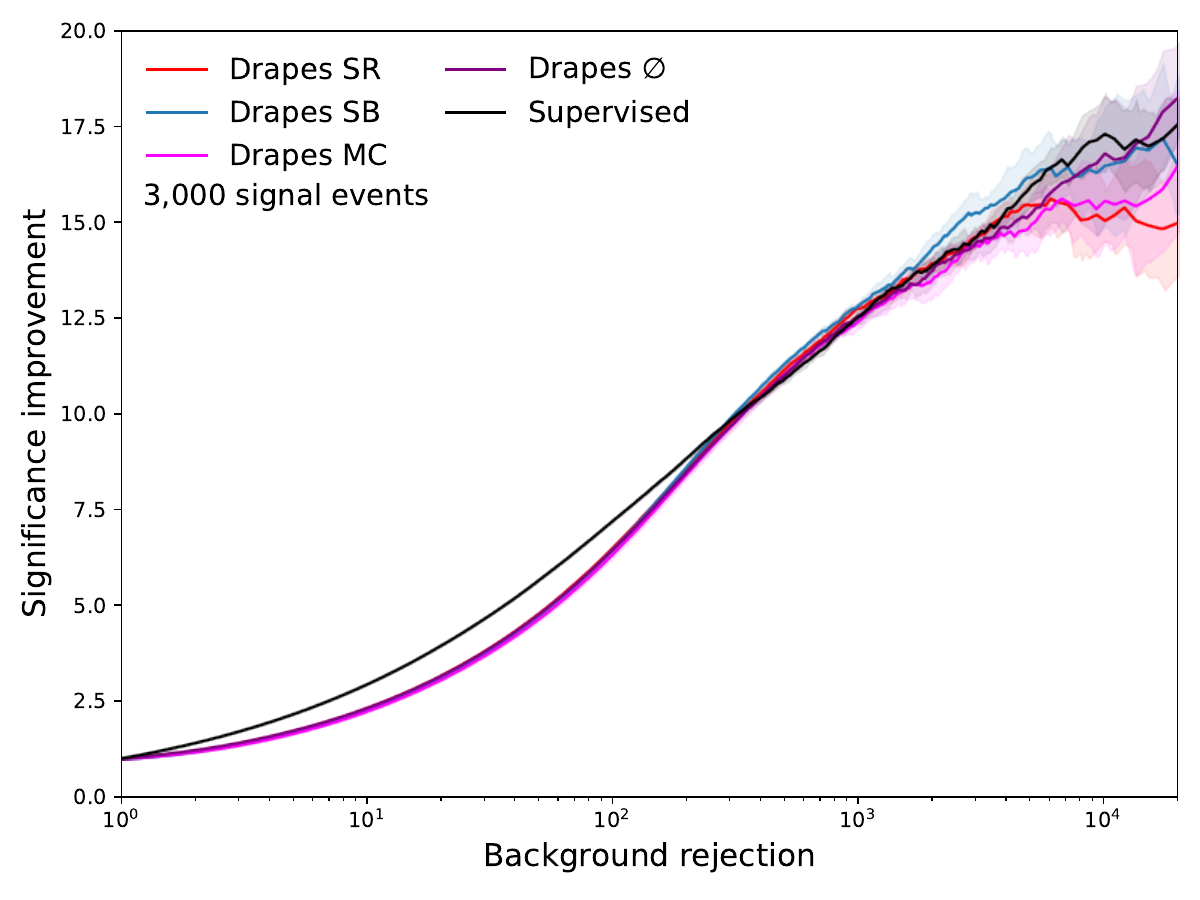}
    \end{subfigure}
    \caption{Background rejection as a function of signal efficiency (left) and significance improvement as a function of background rejection (right)
    for \drapesGen~(purple), \drapesSR~(blue), \drapesSB~(red), \drapesMC~(magenta), and Supervised~(black).
    \drapesSR, \drapesSB, and \drapesMC use initial samples with a noise rate of $0.6$, and \drapesGen generates samples purely from noise.
    All methods are trained on the sample with 1,000 (top) and 3,000 (bottom) injected signal events, and a signal region $3300\leq\mjj<3700$~GeV.
    The lines show the mean value of fifty independent classifiers, with the shaded band representing a 68\% uncertainty.
    }
    \label{fig:sic_roc_extra}
\end{figure}

The significance improvement and receiver operator characteristic curves comparing \drapesGen to the partial diffusion methods are shown in \cref{fig:sic_roc_extra}.
Here a value of $\sigma^\prime=2.24$ is used for all three partial diffusion methods.
At 1,000 injected signal events, \drapesSB is able to reach the highest level of performance.
However, for the same oversampling factor this results in more events in the background reference sample, as each side-band event is considered when generating the background template.
When using the same number of raw training data, \drapesGen achieves even higher levels of performance, as evidenced by the oversampling study in \cref{fig:sic_vs_sig_oversample}, where \drapesGen can reach a max significance improvement of close to 15.

% \clearpage
\subsection{Maximum significance improvement}

\begin{figure}[hbpt]
    \centering
    \begin{subfigure}{0.48\textwidth}
        \centering
        \includegraphics[width=\textwidth]{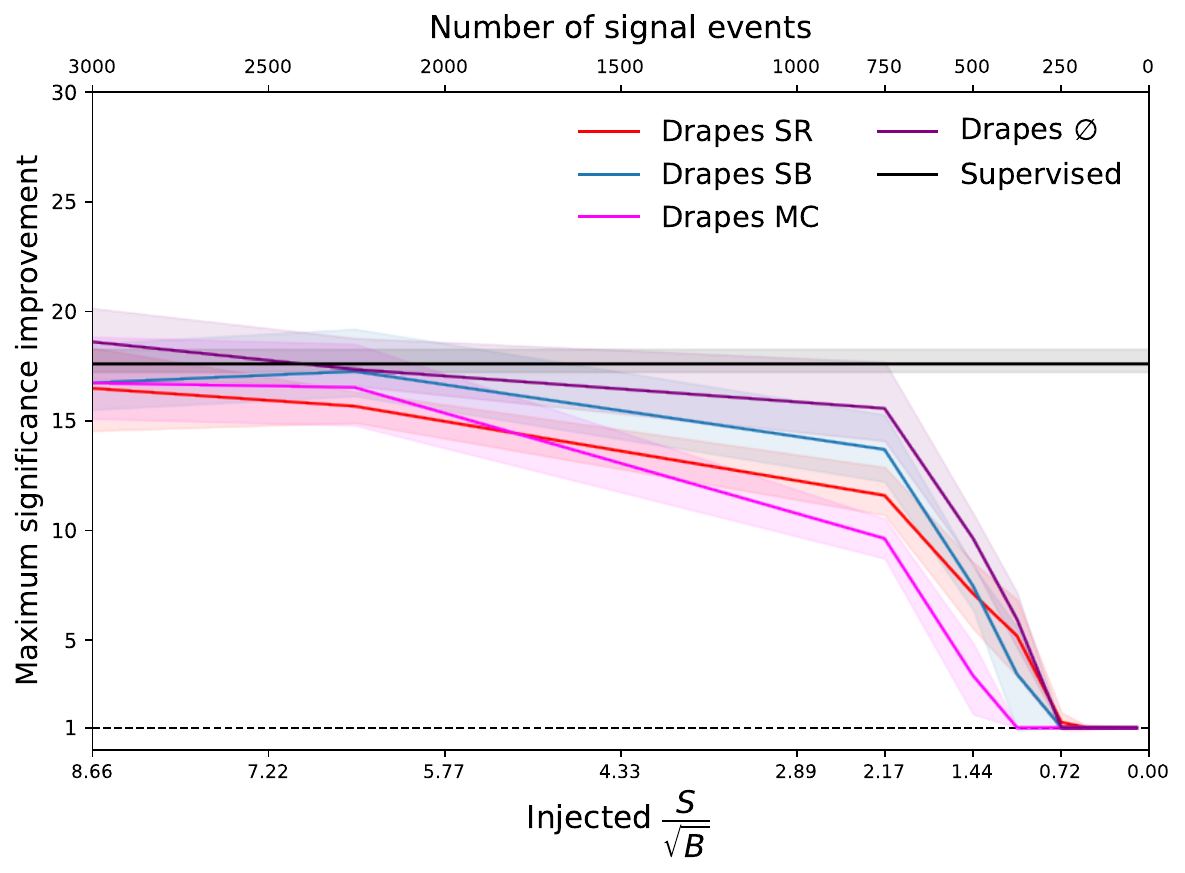}
    \end{subfigure}
    \begin{subfigure}{0.48\textwidth}
        \centering
        \includegraphics[width=\textwidth]{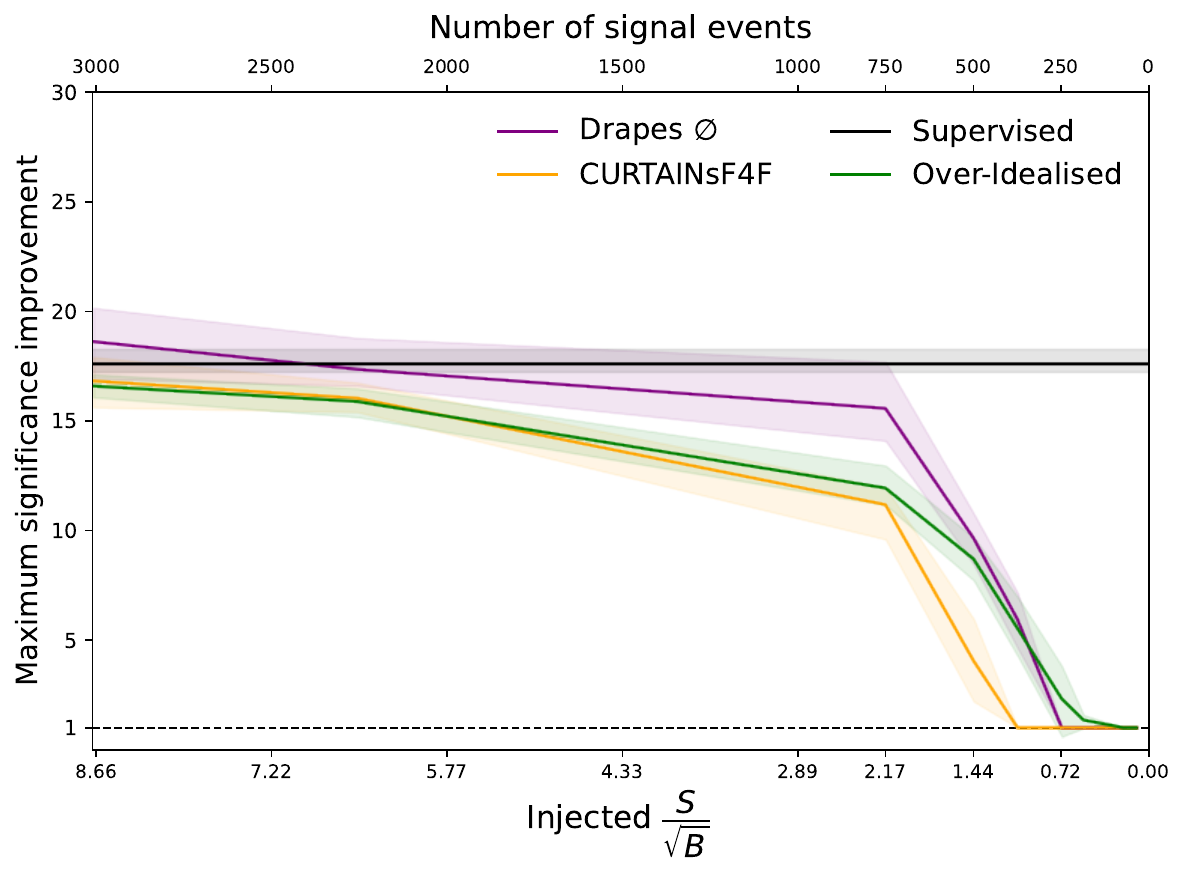}
    \end{subfigure}
    \caption{Maximum significance improvement as a function of the number of signal events in the signal region, \mbox{$3300\leq\mjj<3700$~GeV}, for \FfF~(orange), \drapesGen~(purple), \drapesSR~(red), \drapesSB~(blue), \drapesMC~(magenta), Supervised~(black), and Over-Idealised~(green).
    The lines show the mean value of fifty independent classifiers, with the shaded band representing a 68\% uncertainty.
    }
    \label{fig:sic_vs_sig_all_max}
\end{figure}

\begin{figure}[htbp]
    \begin{subfigure}{0.48\textwidth}
        \centering
        \includegraphics[width=\textwidth]{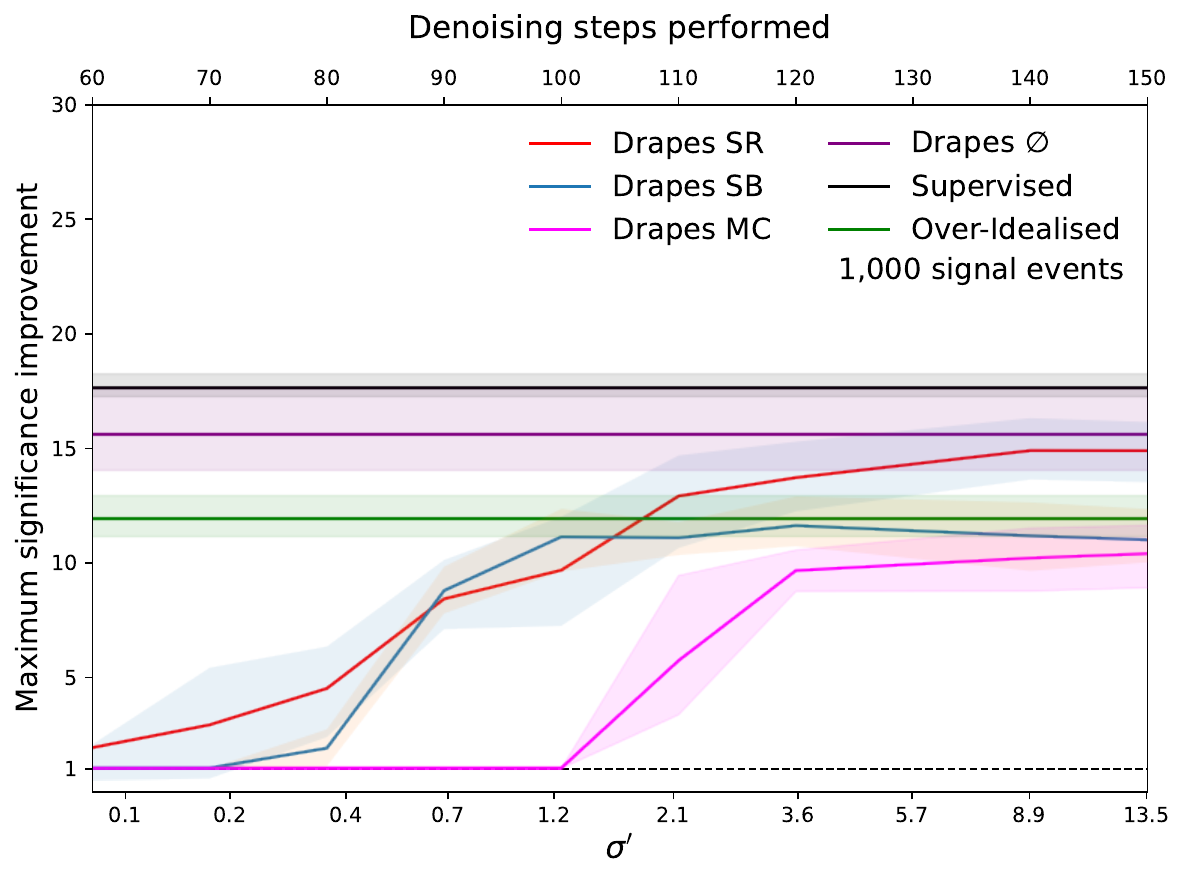}
    \end{subfigure}
    \hfill
    \begin{subfigure}{0.48\textwidth}
        \centering
        \includegraphics[width=\textwidth]{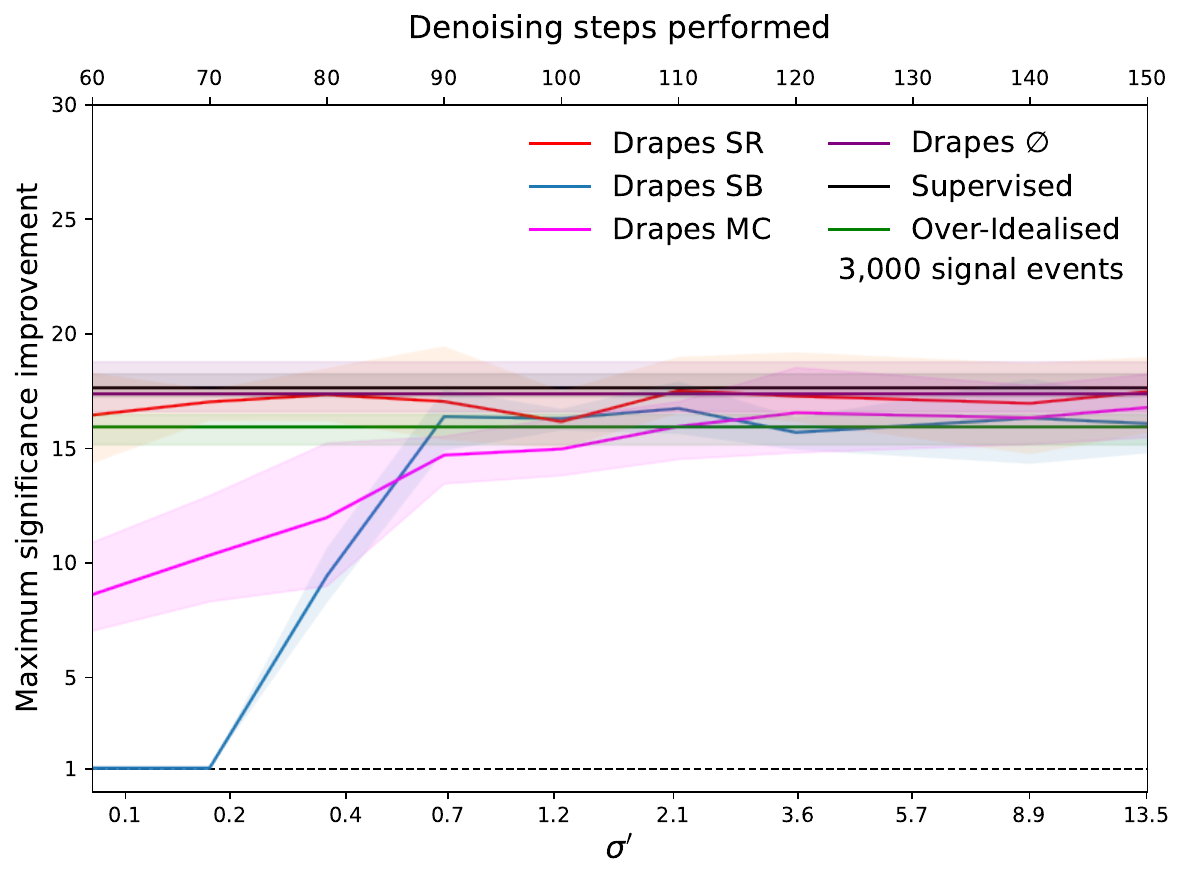}
    \end{subfigure}
    \caption{Maximum significance improvement as a function of noise rate for \drapesSR~(red), \drapesSB~(blue), and \drapesMC~(magenta).
    All generation methods use the same diffusion model trained on the sample with 1,000 (left) and 3,000 (right) injected signal events, and a signal region $3300\leq\mjj<3700$~GeV.
    \FfF~(orange), Supervised~(black) and Over-Idealised~(green) are shown for reference. 
    % A noise rate of 0 represent the input data without any noise perturbation or diffusion steps, and a rate of 1 corresponds to the case where the inputs are completely noise.
    % In the case of \drapesSB target \mjj values for each sample are drawn from a polynomial fit, whereas in \drapesSR the original value for \mjj is used.
    The lines show the mean value of fifty independent classifiers, with the shaded band representing a 68\% uncertainty.
    }
    \label{fig:sic_noise_max}
\end{figure}

In other studies, the maximum significance improvement is used as a measure of performance, despite its sensitivity to statistical fluctuations at very high background rejection.
The maximum significance improvement as a function of the number of injected signal events for all \drapes methods is shown in \cref{fig:sic_vs_sig_all_max}.
The maximum significance improvement as a function of $\sigma^\prime$ for \drapes with partial diffusion is shown in \cref{fig:sic_noise_max}.

% \begin{figure}[htbp]
%     \begin{subfigure}{0.48\textwidth}
%         \centering
%         \includegraphics[width=\textwidth]{figures/solver_sic_vs_mixrate_1000.0.pdf}
%     \end{subfigure}
%     \hfill
%     \begin{subfigure}{0.48\textwidth}
%         \centering
%         \includegraphics[width=\textwidth]{figures/solver_sic_vs_mixrate_5000.0.pdf}
%     \end{subfigure}
%     \caption{Significance improvement as a function of noise rate for \drapesSB when using either the Heun (dashed lines) or Euler (solid lines) solver as a function of the initial noise rate.
%     Both solvers are studied for a maximum number of diffusion steps}
%     \label{fig:solver_noise}
% \end{figure}

\end{document}